\documentclass[twocolumn,aps,showpacs,prb,tightenlines,amsmath,amssymb]{revtex4}
\usepackage{bm}
\usepackage{graphicx}              
\usepackage{amssymb}               
\usepackage{amsmath}                     
\usepackage{colordvi}
\usepackage{calrsfs} 
\usepackage[thinlines,thiklines]{easybmat}
\newcommand{\bgreek}[1]{\mbox{\boldmath$#1$\unboldmath}}
\begin{document}   

\title{A novel superconducting-velocity--tunable quasiparticle state and spin relaxation
  in GaAs (100) quantum wells in proximity to $s$-wave superconductor}
 
\author{T. Yu}
\author{M. W. Wu}
\thanks{Author to whom correspondence should be addressed}
\email{mwwu@ustc.edu.cn.}
\affiliation{Hefei National Laboratory for Physical Sciences at Microscale, Department of Physics,
and CAS Key Laboratory of Strongly-Coupled Quantum Matter Physics,
University of Science and Technology of China, Hefei, Anhui, 230026,
China}

\date{\today}

\begin{abstract} 
  We present a novel quasiparticle state driven by a supercurrent in GaAs (100) quantum wells in
  proximity to an $s$-wave superconductor, which can be tuned by the superconducting
  velocity.  Rich features such as the suppressed Cooper
  pairings, large quasiparticle
  density and non-monotonically tunable momentum current can be realized by
  varying the superconducting
  velocity. In the degenerate regime,
  the quasiparticle Fermi surface is composed by two arcs, referred to as Fermi arcs, which
  are contributed by the electron- and
  hole-like branches. The D'yakonov-Perel' spin relaxation is explored, and intriguing physics is revealed
  when the Fermi arc emerges. Specifically,  
  when the order parameter tends to zero, it is found that the branch-mixing
  scattering is forbidden in the quasi-electron band.
  When the condensation process associated with the annihilation of the
    quasi-electron and quasi-hole is {\em slow}, this indicates that the electron- and hole-like Fermi 
  arcs in the quasi-electron band are independent.
  The open structure of the Fermi arc leads to the nonzero angular-average
  of the effective magnetic field due to the spin-orbit coupling, which acts as
  an effective Zeeman field. This Zeeman field leads to the spin oscillations
  even in the strong scattering regime. Moreover, in the strong scattering
  regime, we show that the open
  structure of the Fermi arc also leads to the insensitiveness of the spin
  relaxation to the momentum scattering, in contrast to the conventional
  motional narrowing 
  situation.  Nevertheless, with a {\em finite} order parameter, the
  branch-mixing scattering can be triggered,
  opening the inter-branch spin relaxation channel, which is
  dominant in the strong scattering regime. In contrast to the situation with
  an extremely small order parameter, due to the inter-branch channel, the spin
  oscillations vanish and the spin relaxation exhibits motional narrowing feature
  in the strong scattering regime.

\end{abstract}

\maketitle
\section{Introduction}
In recent years, the superconducting spintronics has attracted much attention
for providing new methods to control over the spin degree of freedom based
on the spin-triplet Cooper pairs and Bogoliubov
quasiparticles.\cite{Supercurrents,Superconducting_spintronics,Eschrig_supercurrent}
On one hand, the triplet Cooper pairs 
combine both the features of the spintronics\cite{OptOri,Awschalom,Zutic,
  Fabian,Dyakonov_book,wureview,Korn,notebook} and
superconductivity, offering the possibility to realize the spin-polarized
supercurrent.\cite{Supercurrents,Superconducting_spintronics,Eschrig_supercurrent,
  Triplet_S_F,Buzdin_S_F,Tokatly_PRLB,Gorkov_Rashba,Tao,
  semiconductor}
On the other hand, as the quasiparticle charge depends on
its momentum in the conventional superconductor, which is exactly zero at the Fermi
momentum, it is promising to tune the spin degree of
freedom with weak disturbance on the charge one
in one system.\cite{Hershfield,spin_r_t_super,spin_transport,spin_injection1,
  spin_injection_Al,spin_imbalance,spin_transport2,spin_injection2,Takahashi_SHE1,
  Takahashi_SHE2,Hirashima_SHE,Takahashi_SHE3,Takahashi_SHE_exp,injection_Al_SP,Takahashi_SFS}
To reveal the physics in the
superconducting spintronics, the spin dynamics for both the spin-polarized Cooper pairs and quasiparticles
has been intensively
studied.\cite{Supercurrents,Superconducting_spintronics,Eschrig_supercurrent,Hershfield,spin_r_t_super,
spin_transport,spin_injection1,
  spin_injection_Al,spin_imbalance,spin_transport2,spin_injection2,Takahashi_SHE1,
  Takahashi_SHE2,Hirashima_SHE,Takahashi_SHE3,Takahashi_SHE_exp,injection_Al_SP,Takahashi_SFS}

Specifically, for the quasiparticle, rich physics has been reported in the
studies on the charge or spin injection
from the non-magnetic metal or ferromagnet to the conventional
superconductor.\cite{spin_transport,spin_injection1,spin_injection_Al,
  spin_imbalance,spin_injection2,injection_Al_SP,Takahashi_SHE1,
  Takahashi_SHE2,Hirashima_SHE,Takahashi_SHE3,Takahashi_SHE_exp,Takahashi_SFS,Hershfield} 
It is shown that the injection of one electron with charge $e$ into the superconductor can add one
Cooper pair with charge $2e \tilde{v}_{\bf k}^2$ and spin $0$ and a quasiparticle with
charge $e(\tilde{u}_{\bf k}^2-\tilde{v}_{\bf k}^2)$ and spin $1/2$,
respectively.\cite{Takahashi_SHE1,Hirashima_SHE,
  Takahashi_SHE3,spin_injection_Al,Takahashi_SHE_exp,Takahashi_SFS} Here,
$\tilde{u}_{\bf k}^2=1/2+\tilde{\zeta}_{\bf k}/(2E_{\bf k})$
and $\tilde{v}_{\bf k}^2=1/2-\tilde{\zeta}_{\bf k}/(2E_{\bf k})$, in which $\tilde{\zeta}_{\bf
  k}=\varepsilon_{\bf k}-\mu_S$ with $\varepsilon_{\bf k}$ representing the
kinetic energy of the electron and $\mu_S$ being the chemical potential in
  the superconductor;
$E_{\bf k}=\sqrt{\tilde{\zeta}_{\bf k}^2+|\Delta_S|^2}$
denotes the energy spectrum of the quasiparticle with $\Delta_S$ being the
superconducting order parameter. 
Accordingly, in the steady state, the injected charge and spin are
mainly carried by
the Cooper pairs and quasiparticles separately, indicating that the spin-charge
separation can be realized during the injection.\cite{Spin_charge_separation,
  Superconducting_spintronics,spin_imbalance,injection_Al_SP,Takahashi_SHE3}
It is further noticed that in
the process of the charge and spin injections, the non-equilibrium charge and/or spin imbalance can
be created.\cite{Tinkham1,Tinkham2,Takahashi_SFS,spin_imbalance,Takahashi_SHE3,Hershfield}
It is then revealed that in the dynamical process, to maintain the charge
neutrality, the Cooper pair condensate can
respond to the dynamics of the injected
quasiparticles.\cite{Takahashi_SHE1,Takahashi_SHE2,Hirashima_SHE,Takahashi_SHE3,Takahashi_SHE_exp,Hershfield}
Accordingly, the study on the quasiparticle dynamics 
itself is essential to further reveal the dynamics of Cooper pairs.

Among the quasiparticle dynamics, the spin dynamics 
in the superconducting metals has been studied both
theoretically\cite{Hershfield,spin_r_t_super}
and experimentally.\cite{spin_injection_Al,Takahashi_SHE_exp,injection_Al_SP}
Theoretically, the quasiparticle spin relaxation has been
calculated by considering the spin-flip\cite{Hershfield,spin_r_t_super} and
spin-orbit scatterings due to the impurities,\cite{Hershfield} which lies in the Elliott-Yafet 
mechanism.\cite{Yafet,Elliot} In the superconducting state, it is shown that the spin-flip scattering is
efficiently enhanced due to the enhancement of the density of states (DOS).\cite{Hershfield,spin_r_t_super}
Whereas the spin-orbit scattering is efficiently
suppressed due to the coherence factor $\zeta_{\bf k}/E_{\bf k}$ in the
  scattering term.\cite{Hershfield} Experimentally,
the long injection lengths were reported for the spin injected
into the superconducting Al\cite{spin_injection_Al,injection_Al_SP} and Nb,\cite{spin_injection2}
indicating
the long spin relaxation time (SRT) in the superconducting state compared to the
normal one. 
Furthermore, it is further found that the injected spin current in the
superconducting Al can
significantly influences the quasiparticle SRT, with the
  spin relaxation behavior in the superconducting state resembling the normal
one when the injected spin current is large.\cite{spin_injection2} 
As rich physics is revealed in the Elliott-Yafet
mechanism in the superconducting metal,
it is intriguing to study the D'yakonov-Perel' (DP)
mechanism,\cite{DP} which is more important for materials without
center-inversion symmetry, in the superconducting state.    
Furthermore, the proximity-induced
superconductivity has been realized in InAs\cite{2D_super_InAs1,2D_super_InAs2} and GaAs\cite{GaAs28,GaAs29,2D_super_GaAs}
heterostructures, offering the chance to study the DP mechanism in the
superconducting semiconductors.

It is noted that in the study of the dynamics in superconducting system,
 different kinetic equations based on the
 quasiclassical\cite{Gennes1964,Eilenberger_1968,LO_series,Usadel,
Eliashberg,Shelankov_series,Rammer_Smith,Kopnin,Chandrasekhar,Boundary
condition,Eckern,Takane_Japan,Bauerbook,Triplet_S_F,Buzdin_S_F,
Tokatly_PRLB,SU2,SOC_scattering,Unification_Schmid}
  and quasiparticle\cite{Stephen1965,Wolfle0,Wolfle1,Wolfle2,Aronov_full,He3_Combescot,Hara_Nagai,
anomalous_transport,Hershfield,spin_r_t_super,Einzel2008,clean_limit} approximations are used. 
The quasiclassical approximation is applicable for the system with a 
large Fermi energy, in which the dependence on the momentum magnitude is
neglected in the Green function, whereas the frequency and angle-of-momentum
dependencies are explicitly considered.
 For the quasiparticle approximation, in the Green
 function, the dependencies on the angle and
 magnitude of the momentum are explicitly considered, but the
 frequency dependence 
is not emphasized. Moreover, this approximation is applicable only when
 the perturbation on the superconducting order parameter is not strong, 
and hence the quasiparticle energy spectrum is well defined.\cite{Kopnin,Aronov_full,Wolfle0}
 To the best of our knowledge, the quasiparticle 
 approximation is mainly applied to
the system without the 
SOC.\cite{Stephen1965,Wolfle0,Wolfle1,Wolfle2,Aronov_full,He3_Combescot,Hara_Nagai,
anomalous_transport,Hershfield,spin_r_t_super} When there exists the SOC, Einzel
{\em et al.} derived the kinetic equation based on the quasiparticle 
approximation, nevertheless in which the scattering is not considered.\cite{Einzel2008,clean_limit}
 The kinetic equations of quasiparticle with the scattering term explicitly considered in
the presence of the SOC are still absent,
 even for the simplest case with the
$s$-wave order parameter.

In this work, we investigate the DP spin relaxation with 
  superconducting-velocity--tunable quasiparticle state in GaAs (100) quantum wells (QWs) in
  proximity to an $s$-wave superconductor. A novel quasiparticle state is
predicted in the superconducting QWs, 
 based on which the quasiparticle spin relaxation is
then explored. In the $s$-wave superconductor,
the order parameter, i.e., $\Delta_S=|\Delta_S|e^{i\Lambda}$, is contributed by its magnitude
$|\Delta_S|$ and superconducting phase $\Lambda$. 
Then due to the
superconducting proximity
effect,\cite{Dar_Sarma1,Dar_Sarma2,Dar_Sarma3,Hubbard1,Tao} by assuming
  that the superconducting phase is not disrupted by the disorder,
  an $s$-wave order parameter with the same superconducting
phase at the superconductor-semiconductor interface but different
magnitude $|\Delta|$ can arise in the semiconductor. Specifically, with the
inhomogeneous superconducting phase, a superconducting velocity ${\bf
  v}_s=\nabla \Lambda/m^*\equiv {\bf q}/m^*$ arises with $m^*$ being the
 electron effective mass in the QWs.\cite{super_current1,super_current2} Here, it is
assumed that ${\bf v}_s$ is perpendicular to the growth direction of QWs, from which a supercurrent is
induced. In this work,
it is further assumed that the superconducting velocity is small, which marginally influences
the superconducting state in the $s$-wave
superconductor.\cite{super_current1,super_current2}
 However, when $|\Delta|\ll |\Delta_S|$, the superconducting velocity can efficiently tune the
 superconducting state in QWs.

We show that in the superconducting QWs, the superconducting velocity can
cause the tilt of the quasiparticle energy
spectrum. Specifically, in the presence of the supercurrent,
  the energy spectra of the quasi-electron ($+$) and quasi-hole
  ($-$) are
  \begin{equation}
    E^{\pm}_{\bf k}={\bf k}\cdot {\bf
      v}_s/2\pm{\mathcal E}_{{\bf k}}.
    \label{spectrum_e}
  \end{equation}
  Here, ${\mathcal E}_{\bf k}=\sqrt{\epsilon_{\bf k}^2+|\Delta|^2}$
  with $\epsilon_{\bf k}=k^2/(2m^*)+m^*v_s^2/8-\mu$ with $\mu$ being the
    chemical potential in the semiconductor ($\hbar\equiv 1$ throughout this paper). It is noted that
  the chemical potential is shifted by $-m^*v_s^2/8$ due to the superconducting
  velocity. From Eq.~(\ref{spectrum_e}), for the
  quasi-electron band, when $|\Delta|\rightarrow 0$, $E^{+}_{\bf k}\approx ({\bf k}+{\bf
    q}/2)^2/(2m^*)-\mu$ if $|{\bf k}|>k_{\rm F}$ and $E^{+}_{\bf k}\approx -({\bf k}-{\bf
    q}/2)^2/(2m^*)+\mu$ if $|{\bf k}|<k_{\rm F}$. With the former and latter branches
  referred to as the electron- ($\zeta_{\bf k}>0$ with
  negative charge)
  and hole-like ($\zeta_{\bf k}<0$ with positive charge) branches, it can be seen that the superconducting
  velocity leads to the shifts of the electron- and hole-like branches by ${\bf
    q}/2$ and $-{\bf q}/2$, respectively. Then the tilt of the quasiparticle energy
  spectrum can be simply understood.

\begin{figure}[ht]
  {\includegraphics[width=8.5cm]{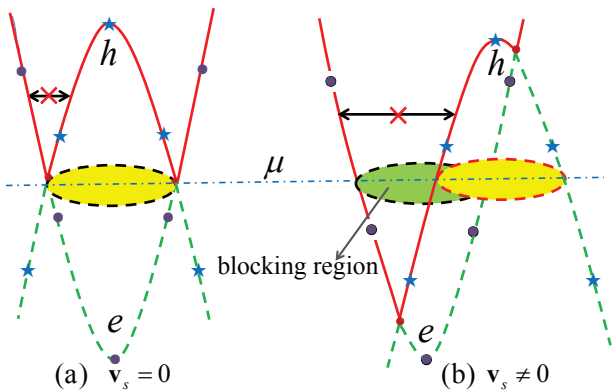}}
  \caption{(Color online) Schematic showing the tilt of the quasiparticle
    energy spectrum and formation of the blocking region. Here, $|\Delta|$ is taken to be
    extremely small. In (a) [(b)], ${\bf v}_s\equiv {\bf q}/m^*=0$ (${\bf
      v}_s\ne0$).
    The red solid (green dashed) curves
    represent the quasi-electron (quasi-hole) energy spectrum; whereas the
    curve labeled by the dots (stars) denotes the electron (hole) band. In (b),
    compared to (a),
    due to the superconducting velocity, the electron and hole-bands are shifted by ${\bf
      q}/2$ and $-{\bf q}/2$, respectively, and hence the resulting quasi-electron
    energy spectrum is tilted . When the quasi-electron energy is
    tilted to be smaller than
    the chemical potential $\mu$ represented by the blue chain line,
    the blocking region emerges, which is represented
    by the green region in the crescent shape in (b). Finally, it is addressed
    that with $|\Delta|$ tending to zero, the branch-mixing scattering
    due to the impurity (represented by the black arrow) is
    forbidden without and with the supercurrent.}
  \label{figyw1}
\end{figure}

  In Fig.~\ref{figyw1}(a) [(b)], the quasi-electron (quasi-hole) energy spectrum
  is schematically plotted with $|\Delta|\rightarrow 0$ when ${\bf v}_s\equiv {\bf q}/m^*=0$ (${\bf v}_s\ne0$) by the red
  solid (green dashed) curves. 
  It can be seen that the quasi-electron (quasi-hole) band is composed by the
  positive (negative) parts of
  the electron and hole bands, shown by the curves labeled by the dots and stars.
  Due to the superconducting velocity, compared to
  Fig.~\ref{figyw1}(a),
  the electron and hole-bands are shifted by ${\bf
    q}/2$ and $-{\bf q}/2$, respectively, and hence the resulting quasiparticle
  energy spectrum is tilted [Fig.~\ref{figyw1}(b)]. Specifically, the tilted
  excitation energy can be even smaller than
  the chemical potential $\mu$, represented by the blue chain line in
  Fig.~\ref{figyw1}.
  Accordingly, the quasi-electrons
  mainly populate at the region with the negative excitation energy even at zero
  temperature, which is referred to
  as the blocking region.\cite{FF,FFLO_Takada}
  In Fig.~\ref{figyw1}(b), the blocking region for the quasi-electron is schematically represented by the green
  ``crescent'', whose formation can be treated as the shift of the Fermi surfaces of
  the electron and hole.
  The appearance of the blocking region can significantly influence
  the Cooper pairings, quasiparticle density and momentum current
  driven by the supercurrent in
  QWs.\cite{super_current1,super_current2,FF,LO,FFLO_Takada}

Specifically, we show that driven by the supercurrent, the
  center-of-mass momentum ${\bf q}$ is carried by the Cooper pairs, with the
  anomalous correlations only existing between the states with momentum ${\bf k}+{\bf q}/2$
  and $-{\bf k}+{\bf q}/2$. Moreover, we show that the anomalous correlations
around the Fermi surface are efficiently suppressed due to the emergence of the
blocking region (refer to
Sec.~\ref{distribution and correlation}). Furthermore,
the quasiparticle density increases with the increase of the superconducting
velocity. In this process, the system 
    experiences the crossover between the non-degenerate and degenerate limits.
Finally, it is revealed that when the blocking region
appears, the momentum current contributed by the quasiparticles flows in the opposite
direction to the one due to the Cooper
pairs. Accordingly, due to the competition of 
the Cooper pairs and quasiparticles in the blocking region, there exists a peak
in the superconducting-velocity dependence of the 
momentum current, whose position 
corresponds to the appearance of the blocking region.

We then study the quasiparticle spin relaxation in the superconducting QWs.
 Based on the quasiparticle approximation due to the small Fermi energy in QWs, 
the kinetic spin Bloch equations (KSBEs)\cite{OptOri,wureview} for the quasiparticle is set
up with the SOC and quasiparticle-impurity
scattering explicitly considered.  By using the KSBEs, we calculate the SRT without and with the
superconducting velocity, respectively. Rich physics is revealed. Without the
supercurrent, we
address that the branch-mixing scattering\cite{Tinkham1,Tinkham2} due to the impurity represented by the black arrow
    in Fig.~\ref{figyw1}(a) is
    forbidden. Here, the
branch-mixing scattering is referred to as the scattering of quasiparticles 
between the electron-like 
 and hole-like
branches.\cite{Tinkham1,Tinkham2} This indicates that the electron- and hole-like
branches are independent and hence only the intra-branch spin relaxation
  channel exists. In this
situation, when $|\Delta|$ tends to zero, the SRT recovers to the normal one.
 Whereas with a finite order parameter, it is found that in the
superconducting state, no matter the scattering is weak or strong, the SRT
is enhanced compared to the normal one, whereas the boundary between the weak and
strong scattering regimes is unchanged. This comes from the efficient
suppressions of the SOC and impurity scattering for the quasiparticle by the same
factor $|\epsilon_{\bf k}|/\mathcal{E}_{\bf k}$.

\begin{figure}[ht]
  {\includegraphics[width=8.7cm]{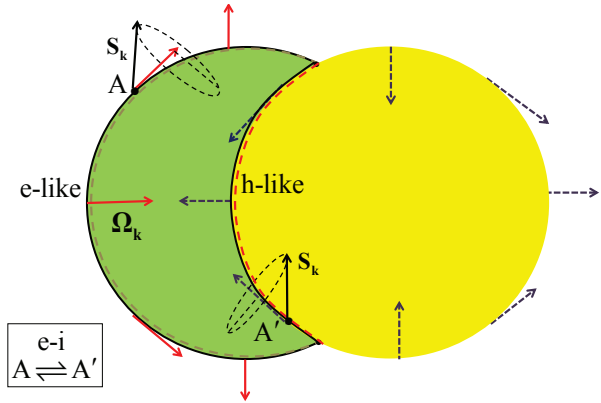}}
  \caption{(Color online) Schematic of the intra- and inter-branch spin relaxation processes.  
    The blocking region is represented
    by the green area in the crescent shape. Around the blocking region, its
    left (right) boundary represented
    by the gray (red) dashed curve mainly comes from the electron(hole)-band,
    which is referred to as the electron(hole)-like Fermi arc in this work. Around
      the electron- and hole-like Fermi arcs, the red solid and blue dashed arrows
      denote the effective magnetic field due to the SOC (i.e., ${\bgreek
        \Omega}_{\bf k}$). With extremely small order parameter, the
    branch-mixing scattering is forbidden, and the spin polarization in the
    electron- and
    hole-like bands relaxes independently. This is referred to as the intra-branch spin
    relaxation.
    Whereas with the finite order parameter, when the blocking region emerges, the
    quasiparticles can be efficiently scattered between the left and right boundaries of the
    blocking region (e.g., scattering from A to A'), triggering the branch-mixing
    scattering.\cite{Tinkham1,Tinkham2} This opens the inter-branch spin relaxation
    channel.}
\label{figyw2}
\end{figure}

With the supercurrent, the quasiparticle spin relaxations with extremely
  small ($|\Delta|\ll 0.1$~meV) and finite ($|\Delta|\gtrsim 0.1$~meV) order
  parameters are explored. 
When the order parameter is extremely small (e.g., $|\Delta|=0.01$~meV), the
branch-mixing scattering is still forbidden [refer to Fig.~\ref{figyw1}(b)]. This is because the coherence factor
($\approx \epsilon_{\bf k}/|\epsilon_{\bf k}|+\epsilon_{{\bf k}'}/|\epsilon_{{\bf k}'}|$) in the
quasiparticle-impurity scattering tends to zero.  However, differing from the situation without the supercurrent, when the
blocking region emerges, the Fermi surfaces from the electron- and hole-like
branches are {\em not closed}, referred to as  ``Fermi
arcs''. In Fig.~\ref{figyw2}, the Fermi arcs from the electron- and
hole-like branches are represented by the gray and red dashed curves in the left and
right boundaries of the blocking region.
One observes that in the electron- or hole-like Fermi arc,
 the angular-average of effective magnetic field due to the SOC (i.e., ${\bgreek
  \Omega}_{\bf k}$) is not zero. When the condensation process is slow,
  which can be associated with the annihilation of quasi-electron and
  quasi-hole,\cite{Bardeen,Josephson,Tinkham_book} the
spin polarizations mainly relaxes within the Fermi arcs.

It is revealed that the quasiparticle spin relaxation at
the Fermi arc {exhibits anomalous features in the {\em strong} scattering
  regime}. Specifically, on one hand, the spin oscillations can be induced by the superconducting
  velocity; on the other hand, the spin relaxation becomes {\em insensitive} to the momentum scattering.
The latter phenomenon is in contrast to the
conventional DP relaxation, where the spin relaxation is suppressed by the
momentum scattering (motional narrowing effect\cite{DP}).
We reveal that the nonzero angular-average of 
  the SOC in one Fermi arc corresponds to an effective Zeeman field. This
  effective Zeeman field can cause the spin oscillations even in the strong
  scattering regime, which nevertheless has little influence on the spin
  relaxation. Actually, this
  feature provides a direct proof for the existence of the Fermi arc.
  It is further shown that by switching off the effective Zeeman
    field, the magnitude of the residue effective magnetic field {\em strongly}
    depends on the direction of the momentum, causing an effective modular-dependent inhomogeneous
  broadening\cite{inhomogeneous,wureview} even for the elastic scattering.
  This modular-dependent inhomogeneous
    broadening can be enhanced by the momentum
  scattering in the
  strong scattering regime and tends to enhance the spin relaxation. Nevertheless, 
  the motional narrowing effect tends to suppress the spin relaxation.\cite{DP} Thus, the competition of the
  two opposite trends leads to the insensitiveness of momentum scattering dependence of the SRT in the
  strong scattering regime.

  When the order parameter is big enough (i.e., $|\Delta| \gtrsim 0.1$~meV in
  our model), in the presence of 
  the supercurrent, it is revealed that the tilt
  of the energy spectrum can trigger the branch-mixing scattering.  
  In this situation,
  there exist the intra- and inter-branch spin relaxation channels for the quasiparticle
  spin relaxation, as illustrated in
  Fig.~\ref{figyw2}. Furthermore, we reveal the role of the intra- and inter-branch spin
  relaxation channels on the spin relaxation in both the weak and strong
  scattering regimes. Specifically, in the weak scattering regime, the intra-branch spin relaxation
  channel is dominant; whereas in the strong scattering regime, the inter-branch
  channel becomes dominant when the blocking region appears. Moreover, in the
  strong scattering regime, with the branch-mixing scattering efficiently
  triggered, the quasi-electron can
  feel all the SOC around the Fermi surface, whose angular-average is
  zero (refer to Fig.~\ref{figyw2}). Accordingly, in the strong scattering regime, in contrast to the situation
  with extremely small order parameter, no spin oscillation
  occurs. Furthermore, in this situation, the inhomogeneous broadening becomes the conventional one
  and hence 
  the spin relaxation is suppressed by the momentum scattering,
  exhibiting the motional narrowing feature.

This paper is organized as follows. We first 
lay out the Hamiltonian in Sec.~\ref{Hamiltonian_Equilibrium}.
In Sec.~\ref{Equilibrium}, we analyze the quasiparticle state in
the superconducting QWs
both analytically (Sec.~\ref{Equilibrium_analytical}) and numerically
(Sec.~\ref{Equilibrium_numerical}).
In Sec.~\ref{KSBE}, the quasiparticle spin relaxation is studied by using the
KSBEs, derived in the quasiparticle
approximation. We conclude and discuss in Sec.~\ref{summary}.

\section{Hamiltonian}
\label{Hamiltonian_Equilibrium}
In this section, we present the Hamiltonian of the symmetric (100) QWs in
proximity to an $s$-wave
superconductor. 
In the particle space, the Hamiltonian is composed by the Bogoliubov-de
Gennes (BdG) Hamiltonian and electron-impurity interaction. The BdG
  Hamiltonian is written as
\cite{Dar_Sarma1,Dar_Sarma2,Dar_Sarma3,Hubbard1,Tao}
\begin{eqnarray}
\nonumber
\hspace{-0.5cm}&&H_{\rm 0}=\int \frac{d{\bf r}}{2}\Phi^{\dagger}({\bf r})
\left(
\begin{array}{cccc}
\zeta_{\bf k} & h_{\bf k} &0&\Delta({\bf r})\\
h^*_{\bf k}&\zeta_{\bf k}&-\Delta({\bf r})&0\\
0&-\Delta^*({\bf r})&-\zeta_{\bf k}&h^*_{\bf k}\\
\Delta^*({\bf r})&0&h_{\bf k}&-\zeta_{\bf k} 
\end{array}
\right)\Phi({\bf r}),\\
\hspace{-0.5cm}
\end{eqnarray}
where $\Phi({\bf r})$ is the particle field operator.
Here, $\zeta_{\bf k}=k^2/(2m^*)-\mu$; $h_{\bf k}=-\alpha
  k_x-i\alpha k_y$ comes from the Dresselhaus SOC,\cite{Dresshaus}
 in which $\alpha=\gamma_D(\pi/a)^2$
  for the infinitely deep well 
  with $\gamma_D$ and $a$ being the Dresselhaus coefficient and well width,
  respectively; $\Delta({\bf
  r})=|\Delta|e^{i{\bf q}\cdot {\bf r}}$ is the $s$-wave order
parameter. Specifically, $|\Delta|$ and ${\bf q}$ are assumed to be 
homogeneous in this work.

The electron-impurity interaction is expressed as 
\begin{eqnarray}
H_{\rm im}=\frac{1}{2}\int d{\bf r}\Phi^{\dagger}({\bf
  r})V({\bf r})\tau_3\Phi({\bf r}),
\end{eqnarray}
with $\tau_3\equiv {\rm diag}(1,1,-1,-1)$ and $V({\bf r})$ denoting the
screened Coulomb potential, whose Fourier component $V_{\bf k}={\displaystyle V^0_{\bf k}}/\big(\displaystyle
    1-P^{(1)}_{\bf k}V^0_{\bf k}\big)$. Here, $V^0_{\bf k}=\displaystyle \int dy\frac{1}{\pi
      a}|I(y)|^2\frac{e^2}{\varepsilon_0\kappa_0(k^2+4y^2/a^2)}$, with   
    $\varepsilon_0$
 and $\kappa_0$ representing the vacuum permittivity and
relative dielectric constant; 
 $|I(y)|^2=\frac{\displaystyle \pi^4\sin^2(y)}{\displaystyle (\pi^2-y^2)^2y^2}$
 standing for the form factor;
$P^{(1)}_{\bf k}$ denoting the longitudinal polarization function, whose
expression has been derived in Ref.~\onlinecite{Tao}.

 In the momentum space,
the BdG Hamiltonian is further represented as
\begin{eqnarray}
\nonumber
\hspace{-0.55cm}&&H_{\rm 0}({\bf k})=\frac{1}{2}\sum_{\bf k}\Phi^{\dagger}_{\bf k}\left(
\begin{array}{cccc}
\zeta_{{\bf k}+\frac{{\bf q}}{2}} & h_{{\bf k}+\frac{{\bf q}}{2}} &
0&|\Delta|\\
h^*_{{\bf k}+\frac{{\bf q}}{2}}&\zeta_{{\bf k}+\frac{{\bf q}}{2}}&-|\Delta|&0\\
0&-|\Delta|&-\zeta_{{\bf k}-\frac{{\bf q}}{2}}&h^{*}_{{\bf k}-\frac{{\bf q}}{2}}\\
|\Delta|&0&h_{{\bf k}-\frac{{\bf q}}{2}}&-\zeta_{{\bf k}-\frac{{\bf q}}{2}}
\end{array}
\right)\Phi_{\bf k},
\end{eqnarray}
where $\Phi^{\dagger}_{\bf k}=\big(a^{\dagger}_{{\bf k}+\frac{{\bf q}}{2}\uparrow},
a^{\dagger}_{{\bf k}+\frac{{\bf q}}{2}\downarrow}, a_{-{\bf k}+\frac{{\bf
      q}}{2}\uparrow},a_{-{\bf k}+\frac{{\bf q}}{2}\downarrow}\big)$; 
the electron-impurity interaction is written as
\begin{eqnarray}
H_{\rm im}=\frac{1}{2}\sum_{{\bf k}{\bf k}'}\Phi_{\bf k}^{\dagger}V_{{\bf k}-{\bf k}'}\tau_3
\Phi_{{\bf k}'}.
\label{impurity}
\end{eqnarray}

We then transform the Hamiltonian in particle space to the 
 quasiparticle one by using the transformation
\begin{equation}
U_{\bf k}=\left(
\begin{array}{cccc}
u_{\bf k} & 0 &0&v_{\bf k}\\
0&u_{\bf k}&-v_{\bf k}&0\\
0&v_{\bf k}&u_{\bf k}&0\\
-v_{\bf k}&0&0&u_{\bf k}
\end{array}
\right).
\label{Unitary}
\end{equation}
Here, 
$u_{\bf k}=\sqrt{\frac{\displaystyle 1}{\displaystyle 2}+\frac{\displaystyle
    \epsilon_{\bf k}}{\displaystyle 2{\mathcal E}_{\bf k}}}$ and
 $v_{\bf k}=\sqrt{\frac{\displaystyle 1}{\displaystyle 2}-\frac{\displaystyle
    \epsilon_{\bf k}}{\displaystyle 2\mathcal{E}_{\bf k}}}$. 
 Then in the quasiparticle space,
the field operator is denoted as 
$\Psi_{\bf k}\equiv (\alpha_{{\bf k}\uparrow}, \alpha_{{\bf k}\downarrow},
\alpha^{\dagger}_{-{\bf k}\uparrow}, \alpha^{\dagger}_{-{\bf k}\downarrow})^T=U_{{\bf k}}\Phi_{\bf k}$.
 Accordingly, the BdG
Hamiltonian in the quasiparticle space is written as 
\begin{eqnarray}
\nonumber
\hspace{-0.43cm}&&H^q_{\rm 0}({\bf k})\\
\nonumber
\hspace{-0.43cm}&&=\left(
\begin{array}{cccc}
{\bf k}\cdot\frac{\displaystyle {\bf v}_s}{\displaystyle 2}+{\mathcal
  E}_{\bf k} &
\frac{\displaystyle \epsilon_{\bf k}}{\displaystyle {\mathcal E}_{\bf k}}h_{\bf
  k}+h_{\frac{\bf q}{2}} &
-\frac{\displaystyle |\Delta|}{\displaystyle {\mathcal E}_{\bf k}}h_{\bf k}&0\\
\frac{\displaystyle \epsilon_{\bf k}}{\displaystyle {\mathcal E}_{\bf k}}h^*_{\bf
  k}+h^*_{\frac{\bf q}{2}}&
{\bf k}\cdot\frac{\displaystyle {\bf v}_s}{\displaystyle 2}+{\mathcal
  E}_{\bf k}&
0&\frac{\displaystyle |\Delta|}{\displaystyle {\mathcal E}_{\bf k}}h^*_{\bf k}\\
-\frac{\displaystyle |\Delta|}{\displaystyle {\mathcal E}_{\bf k}}h^*_{\bf k}&0
&{\bf k}\cdot\frac{\displaystyle {\bf v}_s}{\displaystyle 2}-{\mathcal
  E}_{\bf k}&
\frac{\displaystyle \epsilon_{\bf k}}{\displaystyle {\mathcal E_{\bf k}}}h^*_{\bf
  k}-h^*_{\frac{\bf q}{2}}\\
0&\frac{\displaystyle |\Delta|}{\displaystyle {\mathcal E}_{\bf k}}h_{\bf k}&
\frac{\displaystyle \epsilon_{\bf k}}{\displaystyle {\mathcal E}_{\bf k}}h_{\bf
  k}-h_{\frac{\bf q}{2}}&
{\bf k}\cdot\frac{\displaystyle {\bf v}_s}{\displaystyle 2}-{\mathcal
  E}_{\bf k}
\end{array}
\right).\\
\hspace{-0.43cm}
\label{Hamiltonian_qu}
\end{eqnarray}
The electron-impurity interaction Hamiltonian
is transformed to be 
\begin{equation}
H_{\rm im}
=\frac{1}{2}\sum_{{\bf k}{\bf k}'}\Psi^{\dagger}_{{\bf k}}V^{\rm q}_{{\bf
    k}'-{\bf k}}\Psi_{{\bf k}'},
\label{ei}
\end{equation}
where the impurity potential 
\begin{equation}
V^{\rm q}_{{\bf k}'-{\bf k}}=V_{{\bf k}'-{\bf k}}
\left(
\begin{array}{cccc}
A_{{\bf k}{\bf k}'}  & 0 &0& B_{{\bf k}{\bf k}'} \\
0&A_{{\bf k}{\bf k}'}&-B_{{\bf k}{\bf k}'}&0\\
0&-B_{{\bf k}{\bf k}'}&-A_{{\bf k}{\bf k}'}&0\\
B_{{\bf k}{\bf k}'}&0&0&-A_{{\bf k}{\bf k}'}
\end{array}
\right),
\label{effective}
\end{equation} 
with $A_{{\bf k}{\bf k}'}=u_{\bf k}u_{{\bf k}'}-v_{\bf k}v_{{\bf k}'}$ and
$B_{{\bf k}{\bf k}'}=u_{\bf k}v_{{\bf k}'}+v_{\bf k}u_{{\bf k}'}$.

\section{superconducting-velocity--tunable quasiparticle state in QW${\rm s}$}
\label{Equilibrium}
In this section, we analyze the quasiparticle state in the
superconducting QWs, which can be tuned by the superconducting velocity,
first analytically (Sec.~\ref{Equilibrium_analytical}) and then numerically (Sec.~\ref{Equilibrium_numerical}).

\subsection{Analytical analysis}
\label{Equilibrium_analytical}
In this part, we analytically analyze the quasiparticle state  
by using the equilibrium
Green function at the Matsubara
representation.\cite{Fetter,Mahan,Abrikosov} In the derivation, the SOC is
neglected as it is much weaker compared to the kinetic energy.

In the particle space, the equilibrium Green function at the Matsubara  
representation is defined as
$\tilde{G}_{12}=-i\langle
T_{\tau}\tilde{\Phi}_1\tilde{\Phi}_2^{\dagger}\rangle$,\cite{Fetter,Mahan,Abrikosov}
in which $T_{\tau}$ represents the chronological product; $(1)=(\tau_1,{\bf r}_1)$
is the imaginary$\mbox{-}$time--space point;
$\langle \cdots\rangle$ denotes the ensemble
average;
and $\tilde{\Phi}(t,{\bf r})\equiv e^{i\tau_3\Lambda(t,{\bf r})/2}\Phi(t,{\bf r})$.
The Green function in the frequency-momentum space is derived to be 
\begin{eqnarray}
\nonumber
\hspace{-1cm}&&\tilde{G}(i\omega_n,{\bf k})\\
\hspace{-1cm}&&=\left(
\begin{array}{cccc}
A(i{\omega_n},{\bf k}) & 0 & 0&C(i{\omega_n},{\bf k})\\
0&A(i{\omega_n},{\bf k})&-C(i{\omega_n},{\bf k})&0\\
0&-C(i{\omega_n},{\bf k})&B(i{\omega_n},{\bf k})&0\\
C(i{\omega_n},{\bf k})&0&0&B(i{\omega_n},{\bf k})
\end{array}
\right),
\label{Green_particle}
\end{eqnarray}
where 
\begin{eqnarray}
\left\{
\begin{array}{cccc}A(i{\omega_n},{\bf k})=\frac{\displaystyle i\omega_n+\zeta_{{\bf k}-{\bf
      q}/2}}{\displaystyle (i\omega_n-\zeta_{{\bf
      k}+{\bf q}/2})(i\omega_n+\zeta_{{\bf k}-{\bf q}/2})-|\Delta|^2}\\
B(i{\omega_n},{\bf k})=\frac{\displaystyle i\omega_n-\zeta_{{\bf k}+{\bf
      q}/2}}{\displaystyle (i\omega_n-\zeta_{{\bf
      k}+{\bf q}/2})(i\omega_n+\zeta_{{\bf k}-{\bf q}/2})-|\Delta|^2}\\
C(i{\omega_n},{\bf k})=\frac{\displaystyle |\Delta|}{\displaystyle (i\omega_n-\zeta_{{\bf k}+{\bf
       q}/2})(i\omega_n+\zeta_{{\bf k}-{\bf q}/2})-|\Delta|^2}
\end{array}
\right..
\end{eqnarray}
 Here, 
$\omega_n=(2n+1)\pi k_BT$ are the Matsubara frequencies with $n$ being
  integer and $T$ representing the temperature.
From this Green function, one
obtains the particle density matrix at the equilibrium state,
\begin{eqnarray}
\rho_e^c({\bf k})=\left(
\begin{array}{cccc}
{\mathcal A}({\bf k}) & 0 & 0&{\mathcal C}({\bf k})\\
0&{\mathcal A}({\bf k})&-{\mathcal C}({\bf k})&0\\
0&-{\mathcal C}({\bf k})&{\mathcal B}({\bf k})&0\\
{\mathcal C}({\bf k})&0&0&{\mathcal B}({\bf k})
\end{array}
\right),
\label{rho_particle}
\end{eqnarray}
whose diagonal elements denote the electron and hole distributions, and the
off-diagonal elements represent the anomalous correlations due to the
superconducting order parameter. In Eq.~(\ref{rho_particle}),
\begin{eqnarray}
\left\{
\begin{array}{cccc}
\hspace{-0.98cm}\mathcal{A}({\bf k})\equiv\langle a_{{\bf k}+\frac{\bf
      q}{2}\uparrow}^{\dagger}a_{{\bf k}+\frac{\bf q}{2}\uparrow}\rangle
=u_{\bf k}^2f\big(E^{+}_{\bf k}\big)+v_{\bf k}^2f\big(E^{-}_{\bf
  k}\big)\\
\hspace{-0.52cm}\mathcal{B}({\bf k})\equiv\langle a_{-{\bf k}+\frac{\bf
      q}{2}\uparrow}a_{-{\bf k}+\frac{\bf q}{2}\uparrow}^{\dagger}\rangle=v_{\bf k}^2f\big(E^{+}_{\bf k}\big)+u_{\bf
  k}^2f\big(E^{-}_{\bf k}\big)\\
\mathcal{C}({\bf k})\equiv\langle a_{-{\bf k}+\frac{\bf
      q}{2}\downarrow}a_{{\bf k}+\frac{\bf q}{2}\uparrow}\rangle=u_{\bf k}v_{\bf k}f\big(E^{+}_{\bf k}\big)-u_{\bf k}v_{\bf
  k}f\big(E^{-}_{\bf k}\big)
\end{array}
\right.,
\label{correlations}
\end{eqnarray}
 where $f(E_{\bf k})=1/\{\exp[E_{\bf k}/(k_BT)]+1\}$ is the Fermi
distribution function. 
For the quasiparticle, by a unitary transformation, the density matrix
  at the equilibrium state is 
\begin{eqnarray}
\nonumber
\hspace{-1cm}&&\rho^h_e({\bf k})=U_{\bf k}\rho^c_e({\bf k})U^{\dagger}_{\bf k}\\
\hspace{-1cm}&&={\rm diag}\Big\{f(E^{+}_{\bf k}),f(E^{+}_{\bf k}),
f(E^{-}_{\bf k}),f(E^{-}_{\bf k})\Big\},
\label{quasi_distribution_eq}
\end{eqnarray}
in which only the diagonal elements exist, denoting the quasi-electron and
quasi-hole distributions.

From the quasiparticle distribution at the equilibrium state, one specific
feature arises due to the modification of the energy spectrum by the
superconducting
velocity.\cite{super_current1,super_current2,FF,LO,FFLO_Takada}
It is noted that when ${\bf v}_s=0$, $E^{+}_{\bf k}$ ($E^{-}_{\bf k}$)
is always bigger (smaller)
than zero. When ${\bf v}_s\ne 0$, it can be found that when $m^*v_s^2\mu/2>|\Delta|^2$, there exist
regions in which $E^{+}_{\bf k}<0$ and $E^{-}_{\bf k}>0$ are satisfied. 
These regions are referred to as the blocking region because it is occupied by the
quasi-electrons even at zero
temperature.\cite{FF,LO,FFLO_Takada}
Specifically, the blocking region for
the quasi-electron is written as 
\begin{eqnarray}
\hspace{-0.1cm}\left\{
\begin{array}{c}
\hspace{-0.1cm}-\sqrt{\frac{\displaystyle
    m^*v_s^2\mu}{\displaystyle 2}-|\Delta|^2}<\zeta_{\bf
  k}-\frac{\displaystyle m^*v_s^2}{\displaystyle 8}<\sqrt{\displaystyle
  \frac{m^*v_s^2\mu}{\displaystyle 2}-|\Delta|^2}\\
\hspace{-1.3cm}-1 \le \cos\theta_{\bf k}<-\sqrt{\frac{\displaystyle
  \big[\zeta_{\bf k}+m^*v_s^2/8\big]^2+|\Delta|^2}{\displaystyle m^*v_s^2/2[\zeta_{\bf k}+\mu]}}
\end{array}
\right.,
\label{region_electron}
\end{eqnarray}
with $\theta_{\bf k}$ being the angle between the momentum and superconducting velocity.

Finally, it is addressed that in the Fulde-Ferrell-Larkin-Ovchinnikov
state,\cite{FF,LO,FFLO_Takada}
the Zeeman-field--induced center-of-mass momentum of Cooper pairs plays the similar role as the
superconducting velocity here.

\subsection{Numerical results}
\label{Equilibrium_numerical}
In this part, we present the numerical results for the properties of the
quasiparticle state in GaAs QWs. All parameters including the band structure
and material parameters used in our computation are
listed in Table~\ref{material}.\cite{material_book,parameter_SOC} In the table, $n_e$ is the electron density, and 
$P_0$ represents the initial spin polarization.

\begin{table}[h]
\caption{Parameters including the band structure
and material parameters used in the computation.\cite{material_book,parameter_SOC}}
 \label{material} 
\begin{tabular}{ll|ll}
    \hline
    \hline    
    $m^*/m_0$&\;\;\;\;\;$0.067$&\;$n_e$~(cm$^{-2}$)&\;\;\;\;\;$10^{11}$\\[4pt]
    $\kappa_0$&\;\;\;\;\;$12.9$ &\;$\gamma_D~({\rm eV}\cdot{\rm \mathring{A}}^3)$&\;\;\;\;\;$23.9$\\[4pt]
    $\kappa_{\infty}$&\;\;\;\;\;$10.8$&\;$a~({\rm nm})$&\;\;\;\;\;$20$\\[4pt]
    $P_0$&\;\;\;\;\;$1\%$&\;$T~({\rm K})$&\;\;\;\;\;$1$ \\[4pt] 
    \hline
    \hline
\end{tabular}
\end{table}

\subsubsection{Blocking region}
\label{distribution and correlation}

\begin{figure}[ht]
  {\includegraphics[width=8.2cm]{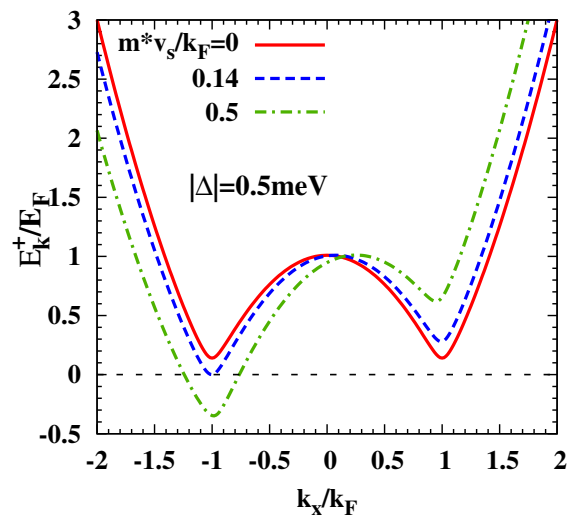}}
  \caption{(Color online) Energy spectra of the quasi-electron with different
    superconducting velocities $m^*v_s/k_{\rm F}=0$ (the red solid curve), $0.14$ (the blue dashed
    curve) and $0.5$ (the green chain curve). $|\Delta|=0.5$~meV in the
    calculation. When $m^*v_s/k_{\rm F}=0$, the
    excitation energy is symmetric for $k_x>0$ and $k_x<0$. However, the finite
    superconducting velocity can cause the tilt of the energy
    spectrum. Specifically, when $m^*v_s/k_{\rm F}\ge 0.14$,
    $E_{{\bf k}}^+\le 0$ can be realized.}
  \label{figyw3}
\end{figure}

We first analyze the energy spectrum of the quasi-electron
[Eq.~(\ref{spectrum_e})]. It is assumed that the superconducting
 velocity is small, which influences
the superconducting state in the $s$-wave
superconductor marginally.\cite{super_current1,super_current2}
 However, when $|\Delta|\ll |\Delta_S|$ here, it can efficiently tune the
 superconducting state in QWs.
Without loss of generality, we assume that the superconducting velocity is along the
$\hat{\bf x}$-direction. In Fig.~\ref{figyw3}, the
$k_x$-dependencies of the energy spectra of the quasi-electron are plotted at different
superconducting velocities.
It can be seen
from the figure that without the supercurrent (the red solid curve), the
excitation energy is symmetric for $k_x>0$ and $k_x<0$. When there is finite
superconducting velocity, the energy spectrum becomes tilted. Specifically, when
$m^*v_s/k_{\rm F}=0.14$, $E_{{\bf k}}^+=0$ can be realized, which is represented by the blue dashed
curve; when $m^*v_s/k_{\rm F}>0.14$, there exist regions with $E_{{\bf k}}^+<0$, as
shown by the green chain curve when $m^*v_s/k_{\rm F}=0.5$. Here, $k_{\rm F}$ is the
Fermi momentum of the electron.

\begin{widetext}
\begin{figure}[htb]
  \begin{minipage}[]{18cm}
    \hspace{0 cm}\parbox[t]{8cm}{
      \includegraphics[width=7.85cm,height=12.74cm]{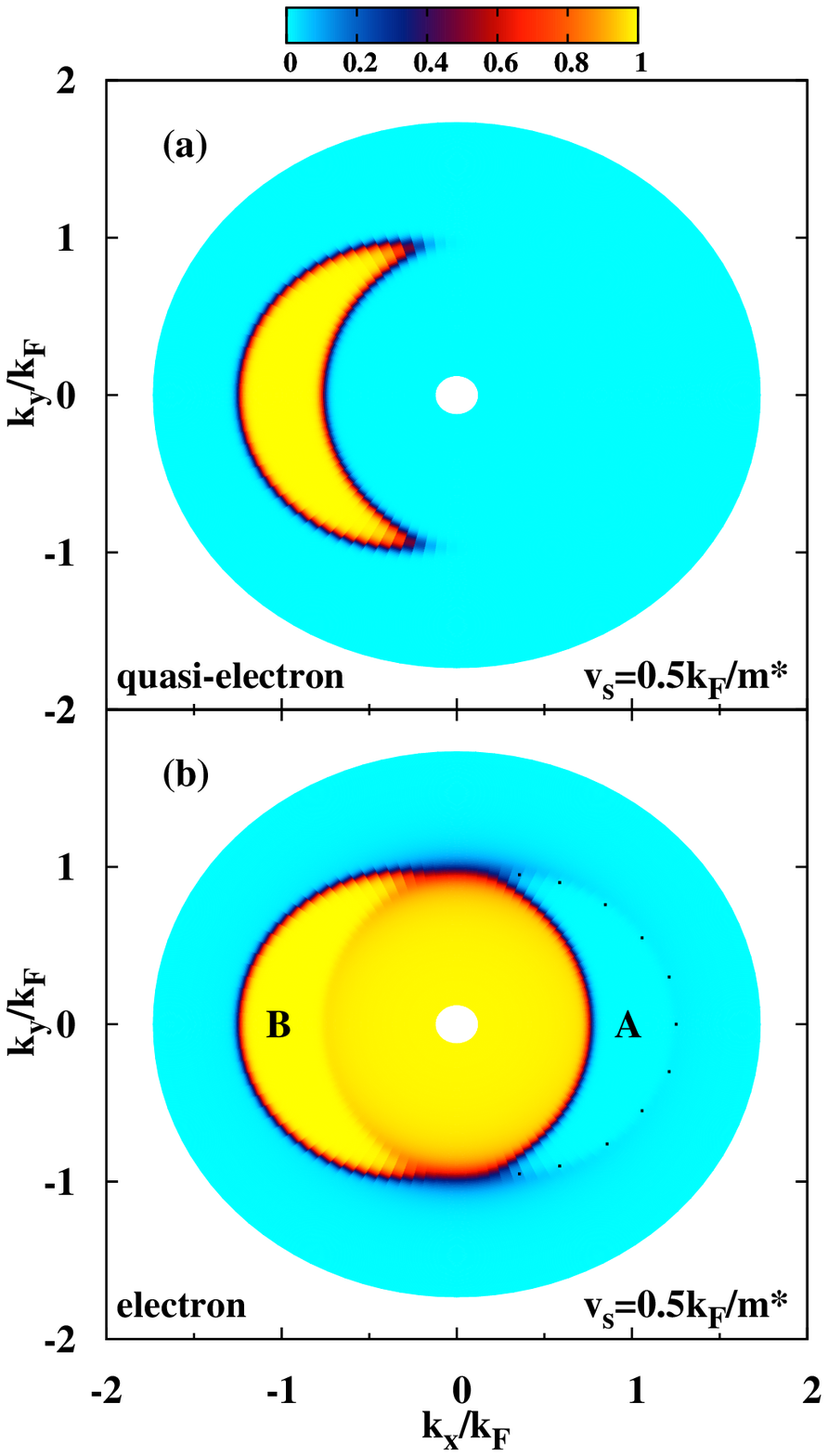}}
    \hspace{0cm}\parbox[t]{8cm}{
      \includegraphics[width=7.85cm,height=12.74cm]{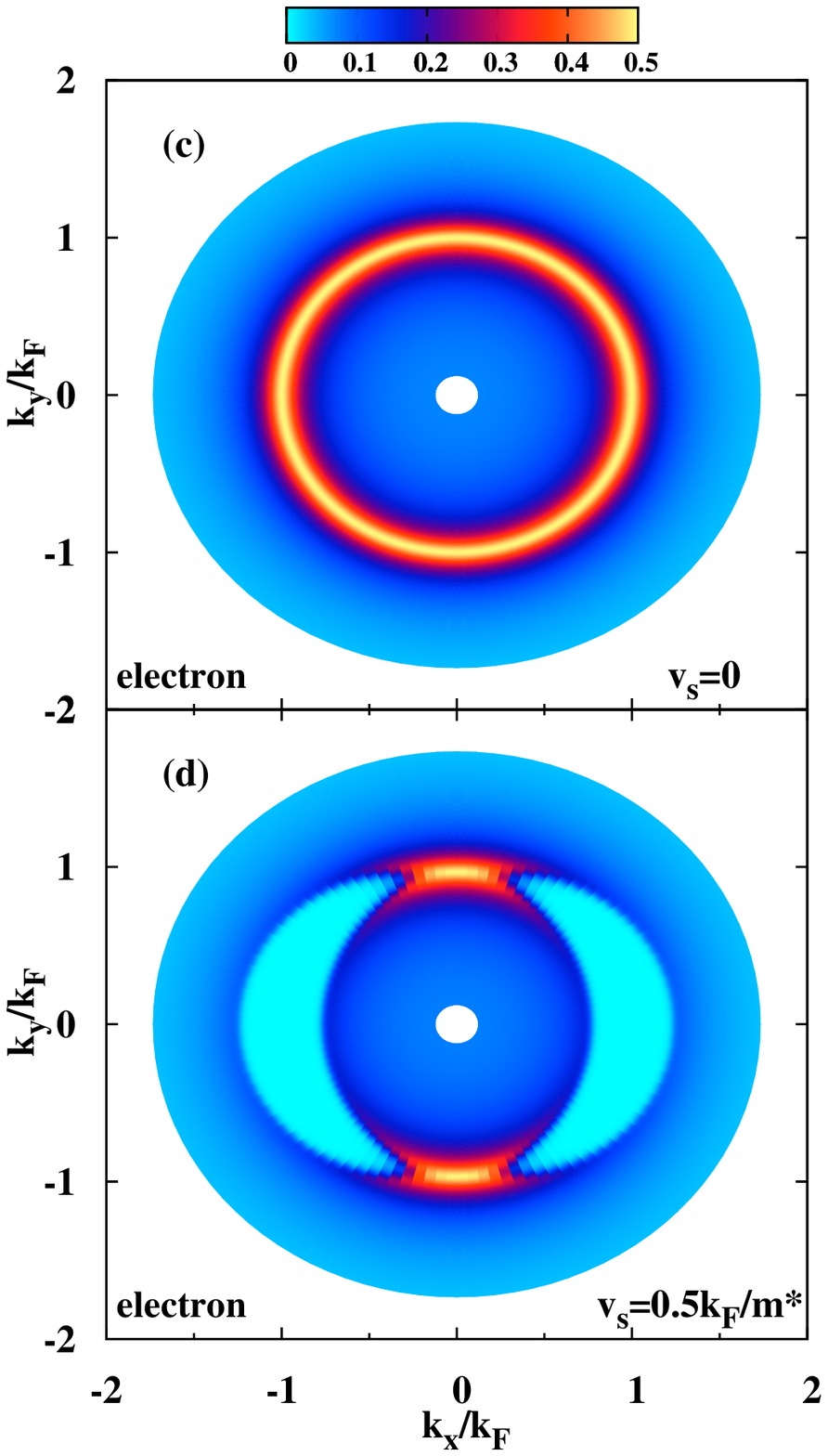}}
  \end{minipage}
\begin{minipage}[]{18cm}
\begin{center}
  \caption{(Color online) Momentum dependencies of the quasi-electron distribution
    [(a)], electron distribution [(b)], and anomalous correlations without [(c)] and
    with [(d)] the supercurrent. (a) is plotted in the quasiparticle space,
    whereas (b), (c) and (d) are shown in the particle space. In (a) and (b), $v_s=0.5k_{\rm
      F}/m^*$. $|\Delta|=0.5$~meV in all the calculation. Specifically,   
    in (a), the quasi-electron distribution is addressed, in which the blocking
    region in the crescent shape arises, whose boundary
    constitutes the ``Fermi surface''. In (b), for the electron
    distribution, it is observed that a crescent
    region when $k_x>0$ (labeled by ``$A$'', enclosed by the dots) disappears and a new crescent
    region (labeled by ``$B$'') appears when
    $k_x<0$. In (c), 
    we show the anomalous correlation without the supercurrent, in which all the electrons around the
    Fermi surface are paired. Finally, in (d), the anomalous correlation with the
    supercurrent ($v_s=0.5k_{\rm
      F}/m^*$) is presented. It can
    be seen that compared to (c), the regions with efficient anomalous
    correlations
    are suppressed to be very small.}
  \label{figyw4}
\end{center}
\end{minipage}
\end{figure}
\end{widetext}

When $E_{{\bf k}}^+<0$, the blocking region appears, which is
determined by Eq.~(\ref{region_electron}). When $v_s=0.5k_{\rm F}/m^*$, the 
quasi-electron population at the equilibrium state is calculated
[Eq.~(\ref{quasi_distribution_eq})], whose momentum dependence is plotted in
Fig.~\ref{figyw4}(a).
It can be seen from Fig.~\ref{figyw4}(a) that the
blocking region is in the crescent shape, whose boundary constitutes the Fermi surface
for the quasi-electron. Furthermore, it can be seen that in the Fermi surface, its left and right
boundaries are contributed by the electron- and hole-like branches,
respectively, in the shapes of arcs, referred to as Fermi arcs in the following.
 It is addressed that even though
  $|\Delta|$ is taken to the finite here, the basic pictures of the blocking
  region as well as the Fermi arcs remain the same as those revealed in Figs.~\ref{figyw1} and \ref{figyw2}.

Actually, the electron distribution in the particle space is also
significantly influenced due to the superconducting velocity [Eq.~(\ref{rho_particle})],
whose Fermi surface is no longer a circle, as shown in
Fig.~\ref{figyw4}(b). It can be seen from Fig.~\ref{figyw4}(b) that a crescent
region at $k_x>0$ (labeled by ``$A$'', enclosed by the dots) disappears and a new crescent region
(labeled by ``$B$'') emerges at $k_x<0$.
The modification of the electron distribution inevitably influences 
the anomalous correlations when there exists a
 supercurrent.
 In the presence of the supercurrent, there exist the anomalous
 correlations between the electron states with momentum 
 ${\bf k}+{\bf q}/2$ and $-{\bf k}+{\bf q}/2$, which can be calculated
 from $\mathcal {C}({\bf k})$ in Eq.~(\ref{correlations}). This indicates that
 the Cooper pairs carry the center-of-mass momentum
 ${\bf q}$, driven by the supercurrent.\cite{super_current1,super_current2}
 From the analysis of the electron distribution in Fig.~\ref{figyw4}(b),
 one finds that the absence of the crescent
 region at $k_x>0$ makes the electrons in the newly arising crescent
 region at $k_x<0$ be unable to find their partners to constitute the Cooper pairs. 
 Accordingly, there is no anomalous correlation for the electrons in the newly arising crescent
 region with $k_x<0$ at zero temperature. The momentum dependence of the
 anomalous correlations 
 without and with the superconducting velocity are explicitly shown in
 Figs.~\ref{figyw4}(c) and (d).
 It can be seen from those two figures that only the electrons around the Fermi
surface can have efficient anomalous correlations. Specifically, in 
Fig.~\ref{figyw4}(c) without the supercurrent, all the electrons around the
Fermi surface are paired. However, when $v_s=0.5k_{\rm
  F}/m^*$ in Fig.~\ref{figyw4}(d), the blocking region appears, in which the anomalous correlation is suppressed to
be close to zero at low temperature. Accordingly, the residue regions with
anomalous correlations are suppressed to be very small when the
  superconducting velocity is large.
This shows that the
superconducting velocity provides an efficient way to tune the Cooper pairing in
the superconducting QWs.

\subsubsection{Quasiparticle density}
\label{BB}
\begin{figure}[ht]
  {\includegraphics[width=8.2cm]{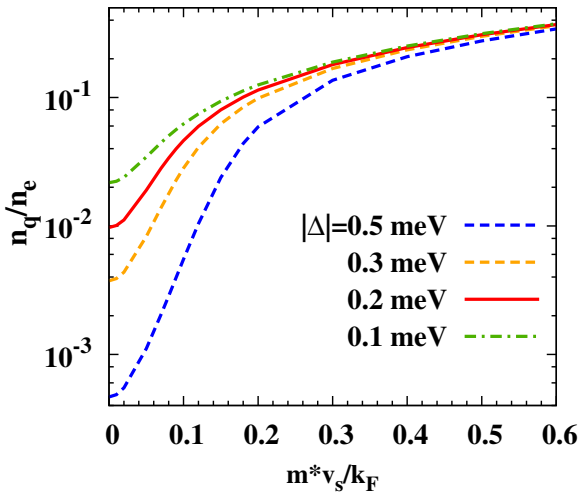}}
  \caption{(Color online) Superconducting velocity dependence of the 
    quasiparticle density with different order parameters $|\Delta|=0.5$~meV (the
    blue dashed curve), 0.3 (the yellow dashed curve), 0.2 (the red solid curve)
    and 0.1~meV (the green chain curve). With the increase
      of the superconducting velocity, the quasiparticle density first increases rapidly and then
    slowly, with the turning point corresponding to the appearance of the
  blocking region approximately.}
  \label{figyw6}
\end{figure}

In this part, we show that the quasiparticle density in QWs can be
efficiently tuned by the superconducting velocity. The quasiparticle density is
calculated from the quasiparticle distributions in the presence of the
superconducting velocity: 
\begin{equation}
  n^{q}=\sum_{\bf k}\big[f(E_{{\bf
      k}\uparrow}^{+})+f(E_{{\bf
      k}\downarrow}^{+})\big].
\end{equation}
In Fig.~\ref{figyw6}, the superconducting velocity dependence of the 
quasiparticle density with different order parameters $|\Delta|=0.5$, 0.3, 0.2
and 0.1~meV are plotted by the blue dashed,
yellow dashed, red solid and green chain curves. It is shown that with the increase of the
superconducting velocity, the quasiparticle density first increases rapidly and then
slowly, with the turning point corresponding to the appearance of the
blocking region roughly. Specifically, with the increase of the
  quasiparticle density due to the superconducting velocity, the blocking
  region and the Fermi surface emerge. In this process, the system can
  experience the crossover between the non-degenerate and degenerate limits.
  Finally, it is noticed that when the superconducting velocity
is large enough, the quasiparticle density is comparable to the
one in the
normal state.

\subsubsection{Momentum current}

\begin{figure}[ht]
  {\includegraphics[width=8.2cm]{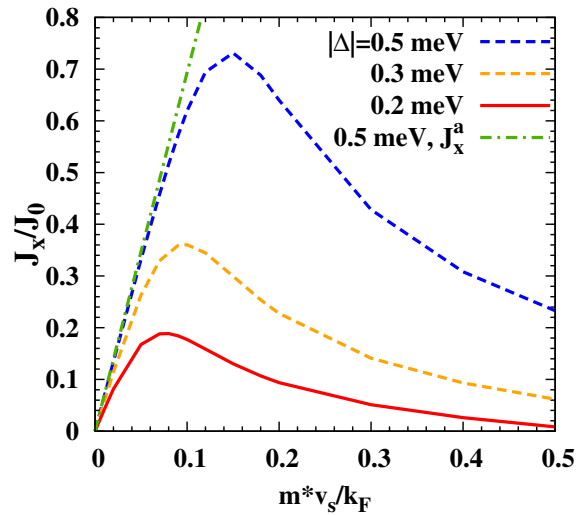}}
  \caption{(Color online) Superconducting velocity dependence of the momentum current with
 different order parameters $|\Delta|=0.5$ (the blue dashed curve),
 0.3 (the yellow dashed curve) and 0.2~meV (the red solid curve).
 The unit for the momentum current is set to be $J_0\equiv
n_ek_{\rm F}(k_BT/E_{\rm F})$. The green chain curve shows the results calculated from
Eq.~(\ref{linear_order}) when $|\Delta|=0.5$~meV, which describes the behavior of momentum current
with small superconducting velocity fairly well.}
  \label{figyw7}
\end{figure}

In the presence of a finite superconducting velocity, the momentum current arises in
the QW, which is calculated
from the equilibrium density matrix, 
\begin{eqnarray}
\nonumber
\hspace{-0.1cm}&&{\bf J}=\sum_{{\bf
    k}}\frac{1}{2}\mbox{Tr}\Big\{\tau_3\big[U_{\bf k}\rho_{e}^h({\bf
    k})U_{\bf k}^{\dagger}+\frac{1}{2}(\tau_3-1)\big]\\
\hspace{-0.1cm}&&\mbox{}\times{\rm diag}\Big({\bf k}+\frac{\bf q}{2},{\bf
  k}+\frac{\bf q}{2},
-{\bf k}+\frac{\bf q}{2},-{\bf k}+\frac{\bf q}{2}\Big)\Big\}.
\label{current}
\end{eqnarray}  
Obviously, ${\bf J}_y=0$ when ${\bf q}=q\hat{\bf x}$. In Fig.~\ref{figyw7}, the
superconducting-velocity dependencies of ${\bf J}_x$ are 
plotted with 
different order parameters $|\Delta|=0.5$, 0.3 and 0.2~meV.
 The unit for the momentum current is set to be $J_0\equiv
n_ek_{\rm F}(k_BT/E_{\rm F})$ with $E_{\rm F}$ being the Fermi energy for the electron. It is shown that 
with the increase of the superconducting velocity, 
the momentum current first increases linearly and then decreases
slowly, with the emergence of a peak. By defining the superconducting velocity corresponding to the
peak of the current as the critical velocity, it can be seen that the critical velocity
increases with the increase of the order parameter.

 We start our analysis from the case with small superconducting velocity (i.e., before the
 appearance of the blocking
 region). With small superconducting velocity satisfying ${\bf k}\cdot {\bf v}_s\ll
 \mathcal{E}_{\bf k}$, the quasiparticle distribution function can be
expanded as a series of ${\bf v}_s$. By keeping ${\bf v}_s$ to its linear
order and considering $|\Delta|\gg k_BT$ here,
 from Eqs.~(\ref{rho_particle}) and (\ref{current}), the momentum current
is calculated to be  
\begin{eqnarray}
\nonumber
{\bf J}^a_x&\approx&m^* v_s\sum_{\bf k}\Big\{u_{\bf k}^2f(\mathcal{E}_{\bf
  k})+v_{\bf k}^2\big[1-f(\mathcal{E}_{\bf k})\big]\Big\}\\
\mbox{}&+&\sum_{\bf k}k_x^2v_s\frac{\partial f(\mathcal{E}_{\bf k})}{\partial
  \mathcal{E}_{\bf k}}\approx m^* v_s\sum_{\bf k}v_{\bf k}^2.
\label{linear_order}
\end{eqnarray} 
It is noted that $v_{\bf k}^2$ is just the distribution function of the Cooper pair
condensate when ${\bf v}_s=0$ (e.g., refer to Takahashi {\em et
    al.}\cite{super_current1,super_current2,Takahashi_SHE3}), 
and hence here the momentum current is mainly carried by the Cooper pairs. The results calculated from
Eq.~(\ref{linear_order}) when $|\Delta|=0.5$~meV is plotted by the green chain curve in
Fig.~\ref{figyw7}, which almost coincides with the blue dashed
curve when ${\bf v}_s$ is small.
However, when the superconducting velocity is large, i.e., with the appearance of the 
blocking region, the current contributed from the quasiparticles becomes
significant. Specifically, the quasi-electron (quasi-hole) mainly populate with
$k_x<0$ ($k_x>0$). Therefore, from
Eq.~(\ref{current}), the current contributed by the quasiparticles
flows in the {\em opposite} direction to the one carried by the Cooper pairs.
 Accordingly, there exists the
competition between the quasiparticles and Cooper pairs, leading to the
critical velocity. Thus, the critical velocity can be estimated by 
$v_s^c\approx \sqrt{2|\Delta|^2/(m^*\mu)}$,
which just corresponds to the appearance of the blocking region.

\section{Quasiparticle spin relaxation}
\label{KSBE}
In this section, we study the quasiparticle spin relaxation in the
superconducting  
GaAs (100) QWs.
 We first derive the KSBEs for the quasiparticle in the
 quasiparticle approximation,\cite{Stephen1965,Wolfle0,
Wolfle1,Wolfle2,Aronov_full,He3_Combescot,Hara_Nagai,
anomalous_transport,Hershfield,spin_r_t_super,Einzel2008,clean_limit} via the
nonequilibrium Green
 function method with the generalized Kadanoff-Baym (GKB) ansatz\cite{Haug,wureview,GKB} (Sec.~\ref{Derivation}), and
 then show the numerical results for the SRTs without (Sec.~\ref{without_superfluid})
 and with the supercurrent (Sec.~\ref{with_superfluid}).

\subsection{KSBEs}
\label{Derivation}

\subsubsection{Derivation with full Hamiltonian}

With a small Fermi energy in the semiconductor QWs,
 the derivation of the KSBEs is based on the quasiparticle
approximation.\cite{Stephen1965,Wolfle0,
Wolfle1,Wolfle2,Aronov_full,He3_Combescot,Hara_Nagai,
anomalous_transport,Hershfield,spin_r_t_super,Einzel2008,clean_limit,Haug}
 In the quasiparticle space, 
the contour-ordered non-equilibrium Green function is defined as\cite{Haug,wureview,Wolfle0} 
\begin{equation}
G_{12}=-i\langle T_c
  \Psi_1\Psi_2^{\dagger} \rangle=\left(
\begin{array}{cc}
G_{12}^{++} & G_{12}^{+-}\\
G_{12}^{-+}&G_{12}^{--} 
\end{array}
\right),
\end{equation}
which is an $8\times 8$ matrix.
 Here, $1=(t_1,{\bf r}_1)$ represents the time-space point; $T_c$ denotes the time contour; and $\Psi(t,{\bf
    r})=\big(\alpha_{\uparrow}(t,{\bf r}),\alpha_{\downarrow}(t,{\bf r}),
\alpha^{\dagger}_{\uparrow}(t,{\bf r}),\alpha^{\dagger}_{\downarrow}(t,{\bf
  r})\big)^T$
 is the quasiparticle field operator. Specifically, 
$G_{12}^{++}=-i\langle T\Psi_1\Psi_2^{\dagger} \rangle$, $G_{12}^{+-}\equiv G_{12}^{<}=i\langle
\Psi_2^{\dagger} \Psi_1\rangle$, $G_{12}^{-+}\equiv G_{12}^{>}=-i\langle \Psi_1\Psi_2^{\dagger}
\rangle$ and $G_{12}^{--}
=-i\langle \tilde{T}\Psi_1\Psi_2^{\dagger} \rangle$ with $T$ and $\tilde{T}$
representing the chronological ordering and anti-chronological ordering,
respectively.

The contour-ordered Green function $G_{12}$ satisfies the Gor'kov's equations\cite{Mahan,Abrikosov}
\begin{eqnarray}
\hspace{-0.8cm}&&[i\stackrel{\rightarrow}{\partial}_{t_1}-{\stackrel{\rightarrow}{H^q_{\rm 0}}}(\hat{\bf
  k}_1)]G_{12}=\delta(1-2)\tilde{\tau}_3+\int_c
d3\tilde{\tau}_3\Sigma_{13}G_{32},
\label{Gorkov_1}\\
\hspace{-0.8cm}&&G_{12}[-i\stackrel{\leftarrow}{\partial}_{t_2}-{\stackrel{\leftarrow}{H^q_{\rm 0}}}(-\hat{\bf
  k}_2)]=\delta(1-2)\tilde{\tau}_3+\int_c d3\Sigma_{13}G_{32}\tilde{\tau}_3.
\label{Gorkov_2}
\end{eqnarray}
Here, $\tilde{\tau}_3={\rm diag}({\rm I}_{4\times 4},-{\rm I}_{4\times 4})$, and
 $\Sigma_{13}$ is the self-energy due to the quasiparticle-impurity
interaction.
Specifically, from Eqs.~(\ref{Gorkov_1}) and (\ref{Gorkov_2}), one obtains the
kinetic equations for $G_{12}^{<}$,
\begin{eqnarray}
\hspace{-1cm}&&\Big[i\frac{\stackrel{\rightarrow}{\partial}}{\partial t_1}-{\stackrel{\rightarrow}{H^q_{\rm 0}}}(\hat{\bf
  k}_1)\Big]G_{12}^{<}=\int d3(\Sigma_{13}^RG_{32}^{<}+\Sigma_{13}^{<}G_{32}^A),
\label{G_1}\\
\hspace{-1cm}&&G_{12}^{<}\Big[-i\frac{\stackrel{\leftarrow}{\partial}}{\partial t_2}-{\stackrel{\leftarrow}{H^q_{\rm 0}}}(-\hat{\bf
  k}_2)\Big]=\int d3(G_{13}^R\Sigma_{32}^{>}+G_{13}^{>}\Sigma_{32}^A),
\label{G_2}
\end{eqnarray}
where ``$R$'' and ``$A$'' label the retarded and advanced Green functions.\cite{Wolfle0,
Wolfle1,Wolfle2,Hara_Nagai,Haug}

By defining $(t,{\bf r})=(t_1-t_2,{\bf r}_1-{\bf r}_2)$ and $({\bf
    R},T)=(t_1+t_2,{\bf r}_1+{\bf r}_2)/2$ and then taking the difference of
  Eqs.~(\ref{G_1}) and (\ref{G_2}), one obtains 
\begin{eqnarray}
\nonumber
\hspace{-1cm}&&i\partial_TG_{12}^{<}-\Big[\stackrel{\rightarrow}{H^q_{\rm 0}}(\hat{\bf
  k}_1)G_{12}^{<}-G_{12}^{<}\stackrel{\leftarrow}{H^q_{\rm 0}}(-\hat{\bf k}_2)\Big]\\
\hspace{-1cm}&&=\int
d3(\Sigma_{13}^RG_{32}^{<}+\Sigma_{13}^{<}G_{32}^A-G_{13}^R\Sigma_{32}^{>}-G_{13}^{>}\Sigma_{32}^A).
\label{difference}
\end{eqnarray}
From Eq.~(\ref{difference}), by using the gradient expansion,\cite{Haug} 
the kinetic equation is derived for the Fourier component
of $G^<_{12}$, i.e., 
$G^{<}({\bf R},T;{\bf k},E)=\int dtd{\bf r}e^{iEt-i{\bf k}\cdot{\bf r}}G^{<}({\bf
R},T;{\bf r},t)$, which is written as
\begin{widetext}
\begin{eqnarray}
\nonumber
\hspace{-0.6cm}&&\int \frac{dE}{2\pi}e^{-iEt}\frac{\partial G^{<}({\bf
  R},T;{\bf k},E)}{\partial T}+i\Big[H_{\rm 0}^q,\int \frac{dE}{2\pi}e^{-iEt}G^<({\bf R},T;{\bf k},E)\Big]
-\frac{1}{2}\Big\{\frac{\partial H_{\rm 0}^q}{\partial{\bf R}},\int 
\frac{dE}{2\pi}e^{-iEt}\frac{\partial G^{<}({\bf
  R},T;{\bf k},E)}{\partial {\bf k}}\Big\}\\
\nonumber
\hspace{-0.6cm}&&\mbox{}+\frac{1}{2}\Big\{\frac{\partial H_{\rm 0}^q}{\partial {\bf k}}
,\int \frac{dE}{2\pi}e^{-iEt}\frac{\partial G^{<}({\bf
  R},T;{\bf k},E)}{\partial {\bf R}}\Big\}=-i\int_{-\infty}^{t_1}
dt_3\big[\Sigma^{>}({\bf R},{\bf k};t_1,t_3)G^{<}({\bf R},{\bf k};t_3,t_1)
+G^{<}({\bf R},{\bf k};t_1,t_3)\\
\hspace{-0.6cm}&&\mbox{}\times \Sigma^{>}({\bf R},{\bf k};t_3,t_1)
-\Sigma^{<}({\bf R},{\bf k};t_1,t_3)G^{>}({\bf R},{\bf k};t_3,t_1)
-G^{>}({\bf R},{\bf k};t_1,t_3)\Sigma^{<}({\bf R},{\bf k};t_3,t_1)\big].
\label{tensor}
\end{eqnarray}
\end{widetext}
Here, $[\ ,\ ]$ and $\{\ ,\ \}$ represent the commutator and anti-commutator,
respectively; $\Sigma^{\stackrel{>}{<}}({\bf R},{\bf k};t_1,t_3)=n_i\sum_{{\bf
    k}'}V_{{\bf k}'-{\bf k}}^{q}G^{\stackrel{>}{<}}({\bf R},{\bf
  k}';t_1,t_3)V_{{\bf k}-{\bf k}'}^{q}$
 with $n_i$ standing for the impurity density.
 

In Eq.~(\ref{tensor}), the full BdG Hamiltonian $H_{\rm 0}^q({\bf
  k})$
[Eq.~(\ref{Hamiltonian_qu})]
is used, from which neither the detailed balance nor the 
quasiparticle number conservation are satisfied. This can be seen as
follows. On one hand, the summation over ${\bf k}$ on the right-hand side of
Eq.~(\ref{tensor}) is not zero due to the matrix form of $V_{{\bf k}-{\bf
    k}'}^{\rm q}$, which indicates the violation of detailed balance. 
On the other hand, from the second term in the
left-hand side of Eq.~(\ref{tensor}), the off-diagonal $2\times 2$ blocks in $H_{\rm 0}^q({\bf
  k})$ can break the quasiparticle number conservation because of the precession
between the quasi-electron and quasi-hole. This is not strange because in $H_{\rm 0}^q({\bf
  k})$, the quasiparticle number operator does not commute with the 
BdG Hamiltonian (also the electron-impurity interaction Hamiltonian) 
because of the terms proportional to $\alpha\alpha S^\dagger$ and
$\alpha^{\dagger}\alpha^{\dagger}S$.\cite{Bardeen,Josephson,Hershfield,Tinkham_book}
Here, $S$ and $S^{\dagger}$ are
the annihilation and creation operators for the Cooper
pairs.\cite{Bardeen,Josephson,Hershfield,Tinkham_book}
Specifically, these terms are related to the
annihilation or creation of two quasiparticles to create or annihilate one
Cooper pair.\cite{Bardeen,Josephson,Hershfield,Tinkham_book}

Nevertheless, it suggests that when the process for the quasiparticle annihilation (creation)
to (from) the Cooper pair condensate is
slow compared to the process under investigation, the part violating the quasiparticle number conservation in the full
Hamiltonian can be neglected. This approximation has been well applied in the
  derivation of the quasiparticle kinetic equation (e.g., Refs.~\onlinecite{Aronov_full,Kopnin,Hershfield}).
  Fortunately, once the Hamiltonian violating the  
quasiparticle number conservation is neglected, the detailed
balance is automatically satisfied. We address that this assumption is reasonable
for the {\it weak} SOC and low impurity density in this investigation.

The violation of the detailed balance and
quasiparticle number conservation can also be understood from the
mathematical point of view. It is noted that the Gor'kov space spanned by
the ``electron'' and ``hole'' bands is larger than the real
physical space. For linear operations on 
  the Green function, no consequence will occur. However, for nonlinear
  operations, unphysical consequence may appear.

\subsubsection{KSBEs with detailed balance and quasiparticle number conservation
  satisfied}
\label{recovered}
From the above analysis, to obtain a self-consistent kinetic equation, 
we neglect the off-diagonal $2\times 2$ blocks in both the
  BdG Hamiltonian [Eq.~(\ref{Hamiltonian_qu})] and quasiparticle-impurity interaction
  Hamiltonian [Eq.~(\ref{ei})]. To further derive the scattering term, the GKB
  ansatz and Markovian 
approximation are used.\cite{Haug,wureview,GKB} 
 Specifically, in the GKB ansatz,\cite{Haug,wureview,GKB}   
\begin{eqnarray}
\nonumber
&&G^{\stackrel{>}{<}}({\bf R},{\bf k};t_1,t_2)=\mp \Big[G^R({\bf R},{\bf
  k};t_1,t_2)\rho^{\stackrel{>}{<}}({\bf R},{\bf k};t_2)\\
&&\mbox{}-\rho^{\stackrel{>}{<}}({\bf R},{\bf k};t_1) G^A({\bf R},{\bf
  k};t_1,t_2)\Big],
\label{GKB}
\end{eqnarray}
where $\rho^{\stackrel{>}{<}}({\bf R},{{\bf k}};t)=-i\int dE/(2\pi)G^{\stackrel{>}{<}}({\bf
  R},{\bf k};t,E).$ 
In Eq.~(\ref{GKB}), $G^R({\bf R},{\bf
  k};t_1,t_2)$ and $G^A({\bf R},{\bf
  k};t_1,t_2)$ are further approximated by their free equilibrium forms, written as
\begin{eqnarray}
&&\hspace{-0.9cm}G^{R}({\bf
  k};t_1,t_2)\approx -i\theta(t_1-t_2)\exp\big[-i(t_1-t_2)H'_{\rm 0}({\bf
  k})\big]\\
&&\hspace{-0.9cm}G^{A}({\bf
  k};t_1,t_2)\approx i\theta(t_2-t_1)\exp\big[-i(t_1-t_2)H'_{\rm 0}({\bf k})\big].
\end{eqnarray}
Here, $\theta(x)$ is the Heaviside step function, and $H'_{\rm 0}({\bf
  k})$ does not include the off-diagonal $2\times 2$ blocks. In the Markovian 
approximation,\cite{Haug,wureview}   
\begin{eqnarray}
\nonumber
&&\rho^{\stackrel{>}{<}}({\bf R},{\bf k};t_3)=\exp\big[iH'_{\rm 0}({\bf
  k})(t_1-t_3)\big]\rho({\bf R},{\bf k};t_1)\\
&&\times\exp\big[-iH'_{\rm 0}({\bf
  k})(t_1-t_3)\big].
\end{eqnarray}
Furthermore, due to the
small SOC, its contribution to the second and third terms in the left and right-hand sides of Eq.~(\ref{tensor}) can be neglected.\cite{wureview}

Finally, by taking $t_1\rightarrow t_2$, i.e., $t\rightarrow 0$, from the $E$-integrated Green function, 
one obtains the KSBEs of the quasiparticle,
\begin{eqnarray}
\nonumber
\hspace{-0.4cm}&&\frac{\partial \rho^h_{\bf k}}{\partial T}+\Big(\frac{{\bf
  v}_s}{2}+\frac{\epsilon_{{\bf
  k}}}{\mathcal{E}_{\bf k}}\frac{{\bf k}}{m^*}\tau_3\Big)\cdot\frac{\partial
\rho^h_{\bf k}}{\partial {\bf R}}+\frac{\epsilon_{{\bf
  k}}}{\mathcal{E}_{\bf k}}\Big[-\frac{\partial \mu({\bf R})}{\partial {\bf
R}}\Big]\tau_3\frac{\partial
\rho^h_{\bf k}}{\partial {\bf k}}\\
\nonumber
\hspace{-0.4cm}&&+i\Big[\left(
\begin{array}{cccc}
 h_{\rm soc}^e({\bf k})& 0\\
0&h_{\rm soc}^h({\bf k})
\end{array}
\right),\rho^h_{\bf k}\Big]=-2\pi n_i\sum_{{\bf
    k}'}|V^{\rm eff}_{{\bf k}-{\bf k}'}|^2\\
\nonumber
\hspace{-0.4cm}&&\mbox{}\times\Big[\delta(E^+_{{\bf k}'}-E^+_{\bf k})\frac{1+\tau_3}{2}\big(\rho_{\bf k}^h
-\rho_{{\bf k}'}^h\big)+\delta(E^-_{{\bf k}'}-E^-_{\bf k})\\
\hspace{-0.4cm}&&\mbox{}\times\frac{1-\tau_3}{2}\big(\rho_{\bf k}^h
-\rho_{{\bf k}'}^h\big)\Big],
\label{KSBE_final}
\end{eqnarray}
where 
\begin{eqnarray}
&&|V^{\rm eff}_{{\bf k}-{\bf k}'}|^2=|V_{{\bf k}-{\bf k}'}|^2(u_{{\bf k}}u_{{\bf
    k}'}-v_{{\bf k}}v_{{\bf k}'})^2,
\label{effective_U}\\
&&h_{\rm soc}^e({\bf k})=
\left(
\begin{array}{cc}
0 & \frac{\displaystyle \epsilon_{\bf k}}{\displaystyle \mathcal{E}_{\bf
    k}}h_{\bf k}+h_{\frac{\bf q}{2}}\\
\frac{\displaystyle \epsilon_{\bf k}}{\displaystyle \mathcal{E}_{\bf k}}h^*_{\bf
  k}+h^*_{\frac{\bf q}{2}}&0
\end{array}
\right),\\
\label{SOC_electron}
&&h_{\rm soc}^h({\bf k})=
\left(
\begin{array}{cc}
0 & \frac{\displaystyle \epsilon_{\bf k}}{\displaystyle \mathcal{E}_{\bf
    k}}h^*_{\bf k}-h^*_{\frac{\bf q}{2}}\\
\frac{\displaystyle \epsilon_{\bf k}}{\displaystyle \mathcal{E}_{\bf k}}h_{\bf
  k}-h_{\frac{\bf q}{2}}&0
\end{array}
\right).
\end{eqnarray}

In this investigation at extremely low temperature, the quasiparticle-quasiparticle and
  quasiparticle-phonon interactions are inefficient and hence
  only the quasiparticle-impurity scattering is presented here.
One notes that when the SOC and ${\bf v}_s$ are taken to be zero, Eq.~(\ref{KSBE_final}) can
recover the traditional Boltzmann-like equation for the
quasiparticle.\cite{Stephen1965,Aronov_full,Kopnin,Hershfield,Takahashi_SHE3}
 Moreover, when $|\Delta|$ and ${\bf v}_s$ are taken to be zero,
Eq.~(\ref{KSBE_final}) can also recover the conventional KSBEs for
electrons.\cite{wureview}

In Eq.~(\ref{KSBE_final}), the second, third and fourth terms correspond to the
diffusion, drift and coherence terms, respectively. Their physical origins are clear from the
point of view that the quasiparticle state is the combination of the electron
and hole ones [Eq.~(\ref{Unitary})].\cite{Takahashi_SHE1,Hirashima_SHE,
Takahashi_SHE3,Hershfield,Aronov_full} 
Specifically, for the diffusion term, one
notes that the group velocities for the quasi-electron and quasi-hole are
$\frac{\displaystyle {\bf v}_s}{\displaystyle 2}+\frac{\displaystyle \epsilon_{{\bf
  k}}}{\displaystyle \mathcal{E}_{\bf k}}\frac{\displaystyle {\bf
k}}{\displaystyle m^*}$ and $\frac{\displaystyle {\bf v}_s}{\displaystyle
2}-\frac{\displaystyle \epsilon_{{\bf
  k}}}{\displaystyle \mathcal{E}_{\bf k}}\frac{\displaystyle {\bf
k}}{\displaystyle m^*}$, respectively, which correspond to the
quasiparticle velocities in Eq.~(\ref{KSBE_final}). For the drift term, it can be
seen that the charges for the
  quasi-electron and quasi-hole are
  $-|e|(u_{\bf k}^2-v_{\bf k}^2)=-|e|{\epsilon_{\bf k}}/{\mathcal{E}_{\bf k}}$ and
  $|e|{\epsilon_{\bf k}}/{\mathcal{E}_{\bf k}}$, respectively, which are just
  responsible for the
  electrical force experienced by the quasiparticle under the electrical 
  field. Finally, for the coherence term, the SOC experienced by the
  quasi-electron [Eq.~(\ref{SOC_electron})] is the combination of the ones
  experienced by the electron 
  and hole, which can be calculated from 
\begin{equation}
h_{\rm soc}^e({\bf k})=u_{\bf k}^2\left(
\begin{array}{cc}
0 & h_{{\bf k}+\frac{\bf q}{2}}\\
h^*_{{\bf k}+\frac{\bf q}{2}}&0
\end{array}
\right)+v_{\bf k}^2\left(
\begin{array}{cc}
0 & -h_{{\bf k}-\frac{\bf q}{2}}\\
-h^*_{{\bf k}-\frac{\bf q}{2}}&0
\end{array}
\right).
\end{equation}

\subsection{Numerical results}
\label{Numerical}
In this part, we study the quasiparticle spin relaxation in the
spacially homogeneous system. By numerically solving the KSBEs [Eq.~(\ref{KSBE_final})], one obtains the SRT from the time
evolution of the spin polarization
\begin{equation}
  {\bf S}(t)=\sum_{\bf k}\frac{1}{4n_q}\mbox{Tr}\Big\{\tau_3\big[\rho_{\bf
    k}^h(t)+\frac{1}{2}(\tau_3-1)\big]{\rm diag}({\bgreek \sigma},-{\bgreek \sigma}) \Big\},
\end{equation}
where ${\bgreek \sigma}$ are the Pauli matrices.  
Here, we focus on the situation that the initial spin polarization of the
quasiparticle is along the 
$\hat{\bf z}$-direction, corresponding to the spin imbalance, which can be realized by the spin
injection with small charge imbalance.\cite{spin_transport,spin_injection1,spin_injection_Al,
  spin_imbalance,spin_injection2,injection_Al_SP,Takahashi_SHE1,
  Takahashi_SHE2,Hirashima_SHE,Takahashi_SHE3,Takahashi_SHE_exp,Takahashi_SFS}
Accordingly, the initial density matrix for the quasiparticle is
written as 
\begin{equation}
  \rho^h_{\bf k}(t=0)={\rm diag}\Big[f(E_{{\bf k}\uparrow}^{+}),f(E_{{\bf
      k}\downarrow}^{+}),f(E_{{\bf k}\uparrow}^{-}),f(E_{{\bf k}\downarrow}^{-})\Big], 
\end{equation} 
where $f(E_{{\bf k}\uparrow}^{\pm})=f(E_{{\bf k}}^{\pm}+\mu_{\uparrow}^q)$ and
$f(E_{{\bf k}\downarrow}^{\pm})=f(E_{{\bf k}}^{\pm}+\mu_{\downarrow}^q)$. Here,
$\mu_{\uparrow}^q$ and $\mu_{\downarrow}^q$ are determined by the density of quasi-electrons
with spin-up $n_{\uparrow}^{q}$ and spin-down $n_{\downarrow}^{q}$, i.e.,
\begin{equation}
  n_{\uparrow,\downarrow}^{q}=\sum_{\bf k}f(E_{{\bf
      k}\uparrow,\downarrow}^{+})=\sum_{\bf k}f(E_{{\bf k}}^{+}+\mu^q_{\uparrow,\downarrow}).
\end{equation}
The parameters for the computation are listed in Table~\ref{material}.

\subsubsection{Quasiparticle spin relaxation without a supercurrent}
\label{without_superfluid}
\begin{figure}[ht]
  {\includegraphics[width=8.5cm]{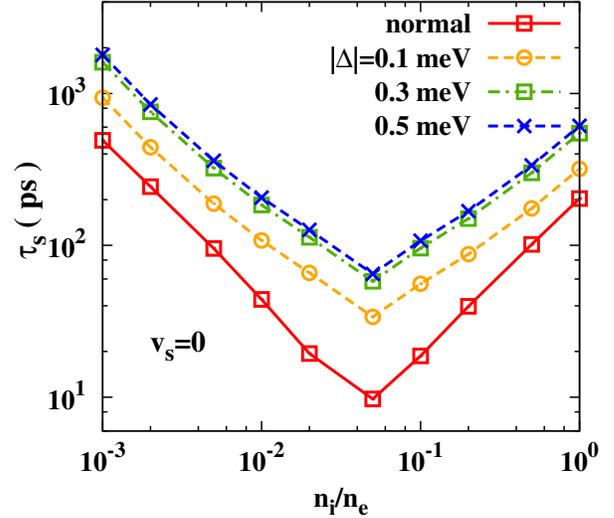}}
  \caption{(Color online) Impurity density dependence of the quasiparticle SRTs
    without the supercurrent when $|\Delta|=0$ (normal state, the red solid
    curve with squares), 0.1 (the yellow dashed
curve with circles), 0.3 (the green chain curve with squares) and 0.5~meV (the blue
dashed curve with crosses). No matter in the weak or strong scattering regime, the SRTs
are enhanced in the superconducting state compared to the normal
one. However, the boundary between the weak and
  strong scattering regimes is unchanged with the variation of the order
  parameter magnitudes.}
  \label{figyw8}
\end{figure}

In this part, we analyze the quasiparticle spin relaxation without the
supercurrent, i.e., ${\bf v}_s=0$,
and reveal the influence of
the magnitude of the order parameter on the quasiparticle spin relaxation. 
When $|\Delta|=0$, the system returns to the normal situation with the
Rashba-like SOC, which has been well studied in GaAs (100) QWs.\cite{wureview}
 In Fig.~\ref{figyw8}, it
is shown that when $n_i\lesssim
0.05n_e$ ($n_i\gtrsim 0.05n_e$), the system lies in the weak (strong)
scattering regime with $\tau_s\propto \tau^N_{\bf k}$ 
[$\tau_s\propto (\tau^N_{\bf k})^{-1}$]. Here, $\tau^N_{\bf k}$ is the momentum
  relaxation time in the normal state. When $|\Delta|$ varies from $0.1$ to $0.5$~meV, the SRTs
  are enhanced in both the weak
  and strong scattering regimes. However, the boundary between the weak and
  strong scattering regimes is unchanged. This can be understood from the following analytical analysis.

From Eq.~(\ref{KSBE_final}), one can see that when ${\bf v}_s=0$, both the
  SOC and momentum scattering for the quasiparticle are modified by the order
  parameter.
 Specifically, from the coherent term [the fourth term in Eq.~(\ref{KSBE_final})], one
  observes that when ${\bf v}_s=0$, the SOC is modified to be
\begin{equation}
h_{\rm soc}^{S}({\bf k})= (\epsilon_{\bf k}/\mathcal{E}_{\bf k})h^N_{\rm soc}({\bf k}),
\label{SOC_eff}
\end{equation}
where $h^N_{\rm
    soc}({\bf k})$ is the normal-state SOC in GaAs (100) QW. Accordingly, due to the
  factor $\epsilon_{\bf k}/\mathcal{E}_{\bf k}$ in Eq.~(\ref{SOC_eff}),
 the SOC is exactly zero when $|{\bf k}|=k_{\rm F}$ and 
efficiently suppressed for the quasiparticles with momentum around
$k_{\rm F}$. For the scattering term, new features arise due to the
modifications of the quasiparticle-impurity interaction potential
[Eq.~(\ref{effective_U})] and the DOS. In the quasiparticle-impurity potential [Eq.~(\ref{effective_U})],
the coherent factor
$(u_{\bf k}u_{{\bf k}'}-v_{\bf k}v_{{\bf k}'})^2$
arises,\cite{Aronov_full,Hershfield} which equals to $(\epsilon_{\bf
  k}/\mathcal{E}_{\bf k})^2$ because in the elastic scattering, the branch
mixing is forbidden\cite{Tinkham1,Tinkham2} and hence $|{\bf
  k}'|=|{\bf k}|$.
 Whereas the DOS is modified to be $(\mathcal{E}_{\bf k}/|\epsilon_{\bf
  k}|)N_0({\bf k})$ with $N_0({\bf k})$ being the DOS in the normal state.
 Accordingly, compared to the normal state, a new prefactor $|\epsilon_{\bf
  k}|/\mathcal{E}_{\bf k}$ arises in the scattering term for the
superconducting state, which is also exactly zero when 
$|{\bf k}|=|{\bf k}'|=k_{\rm F}$. Consequently, the momentum
scattering time in the
superconducting
 state becomes\cite{Hershfield,spin_r_t_super,Bardeen}
\begin{equation}
\tau^{S}_{\bf k}=(\mathcal{E}_{\bf k}/|\epsilon_{\bf
  k}|)\tau_{\bf k}^{N}.
\label{scattering_time} 
\end{equation}
Because at low temperature, the quasiparticles mainly populate around
the states with momentum $|{\bf k}|=k_{\rm F}$, the momentum scattering is
efficiently suppressed in the superconducting state.

Obviously, from Eqs.~(\ref{SOC_eff}) and (\ref{scattering_time}),
 $|h_{\rm soc}^{S}({\bf k})|\tau^{S}_{\bf k}=|h^N_{\rm soc}({\bf
  k})|\tau^{N}_{\bf k}$, and hence the boundary between the weak and strong
scattering regimes remains unchanged when the system enters to the superconducting
state. Then by solving the KSBEs
analytically,
 one obtains the SRTs in both the strong and weak scattering regimes.\cite{wureview,Zhang_graphene,Yu_valley}
 Specifically, in the strong
scattering limit with $\Omega^{S}_{\bf k}\tau^{S}_{\bf k}\ll 1$,
 ${\bf S}_{\bf k}^z(t)\approx P_0\exp[-4(\Omega^S_{\bf k})^2\tau_{\bf k}^S
 t]$. Here,
\begin{equation}
 \Omega^S_{\bf k}=(\alpha k)(\epsilon_{\bf
  k}/\mathcal{E}_{\bf k})=\Omega^N_{\bf k}(\epsilon_{\bf
  k}/\mathcal{E}_{\bf k})
\end{equation}
 is the precession frequency due to the SOC, and
\begin{equation}
\frac{1}{\tau_{\bf k}^S}=\frac{n_im^*}{2\pi}\frac{\epsilon_{\bf k}}{\mathcal{E}_{\bf k}}
\int d\theta_{{\bf k}-{\bf k}'}|V_{{\bf k}-{\bf k}'}|^2(1-\cos\theta_{{\bf k}-{\bf k}'}).  
\end{equation}
In the weak scattering limit with $\Omega^S_{\bf k}\tau_{\bf k}^S\gg 1$,
 ${\bf S}_{\bf k}^z(t)\approx P_0\exp[-t/(2\tau_{\bf k}^S)]\cos(2\Omega_{\bf k}^St)$. 
 On one hand, the momentum scattering opens a spin relaxation 
channel due to the factor $\exp[-t/(2\tau_{\bf k}^S)]$; on the other hand, the factor
$\cos(2\Omega_{\bf k}^St)$ can cause the free induction decay due to different precession
 frequency with different momentum, which is suppressed in the degenerate
 regime.\cite{wureview,Zhang_graphene,Yu_valley}
 Accordingly, the SRT for the quasiparticle with momentum ${\bf k}$ reads  
\begin{eqnarray}
\tau^S_s({\bf k})&\approx&\left\{\begin{array}{cc}
\big[4(\Omega^S_{\bf k})^2\tau_{\bf k}^S\big]^{-1},~~\Omega^{S}_{\bf k}\tau^{S}_{\bf k}\ll 1;\\
2\tau_{\bf k}^S,~~~~~~~~~~~~~~~\Omega^{S}_{\bf k}\tau^{S}_{\bf k}\gg 1
\end{array}\right.\\
&=&(\mathcal{E}_{\bf k}/|\epsilon_{\bf
  k}|)\tau^N_s({\bf k}),
\label{relation}
\end{eqnarray}
with $\tau^N_s({\bf k})$ being the SRT in the normal state. Due
to the factor $\mathcal{E}_{\bf k}/|\epsilon_{\bf k}|$
 in Eq.~(\ref{relation}), no matter in the strong or weak scattering
regime, the SRT for the quasiparticle with the momentum
around $k_{\rm F}$ is {\em enhanced} compared to the normal one.

Finally, based on Eq.~(\ref{relation}), we calculate the total SRT of the
  quasiparticle, written as\cite{wureview,Yu_valley}
\begin{equation}
\frac{1}{\tau_s}=\frac{\displaystyle \sum_{\bf k}\frac{1}{\tau^N_s({\bf k})}\frac{|\epsilon_{\bf
  k}|}{\mathcal{E}_{\bf k}}\big[f(E_{{\bf
    k}\uparrow})-f(E_{{\bf k}\downarrow})\big]}{\displaystyle \sum_{\bf k}\big[f(E_{{\bf
    k}\uparrow})-f(E_{{\bf k}\downarrow})\big]}.
\label{total}
\end{equation}
With the small spin polarization, the spin polarization is limited to the region
around the Fermi surface, and hence in Eq.~(\ref{total}), $\tau^N_s({\bf
  k})\approx \tau^N_s({\bf k}_F)$. Accordingly, from Eq.~(\ref{total}), one
obtains
\begin{equation}
\frac{1}{\tau_s}\approx
\frac{1}{\tau^N_s}\frac{1}{P_0n_q}\frac{m^*}{2\pi}(\mu_{\uparrow}^q-\mu^q_{\downarrow})
\big[f(\sqrt{\mu^2+|\Delta|^2})-2f(|\Delta|)\big].
\end{equation}
Furthermore, when $\mu\gg |\Delta|$ and $\mu^q_{\uparrow}\approx -\mu^q_{\downarrow}\equiv
\delta\mu$ due to
the small spin polarization, one obtains
\begin{equation}
\frac{1}{\tau_s}\approx
\frac{1}{\tau^N_s}\frac{1}{P_0n_q}\frac{2m^*}{\pi}|\delta\mu|f(|\Delta|)=Q\frac{1}{\tau^N_s},
\label{final_relation}
\end{equation}
with $Q\equiv\frac{\displaystyle 1}{\displaystyle P_0n_q}\frac{\displaystyle
  2m^*}{\displaystyle \pi}|\delta\mu|f(|\Delta|)$. Here, with $|\Delta|=0.1$,
 $0.3$ and $0.5$~meV, $1/Q$ is calculated to be $1.7$, $2.7$, and $3.3$, in good
 agreement with the numerical results. From Eq.~(\ref{final_relation}),
this shows that due to the order parameter, the SRT can be enhanced by several
times and increases with the increase of the order parameter.

Finally, it is addressed that we also consider the influence of the {\it inelastic}
quasiparticle-quasiparticle scattering on the quasiparticle spin relaxation,
which is proven to be inefficient even with the
low impurity density $n_i=10^{-3}n_e$.\cite{Coulomb}

\subsubsection{Quasiparticle spin relaxation with a supercurrent}
\label{with_superfluid}

In this part, we consider the influence of the supercurrent or superconducting
velocity on the quasiparticle
spin relaxation with small and large order parameters $|\Delta|=0.01$ and $0.5$~meV, respectively. 
The superconducting velocity can cause the tilt of the quasiparticle energy
spectrum and hence modifies the quasiparticle distribution (refer to 
Sec.~\ref{distribution and correlation}), which is
expected to cause rich physics in the quasiparticle spin
relaxation. It is assumed that the condensation
  process associated with the annihilation of the quasi-electron and quasi-hole
  is slow compared to the spin relaxation rate. Here, we first describe
the rich phenomenon from the impurity density and superconducting
velocity dependencies of the quasiparticle SRTs.

\begin{widetext}
\begin{figure}[htb]
  \begin{minipage}[]{18cm}
    \hspace{0 cm}\parbox[t]{8cm}{
      \includegraphics[width=7.85cm,height=12.74cm]{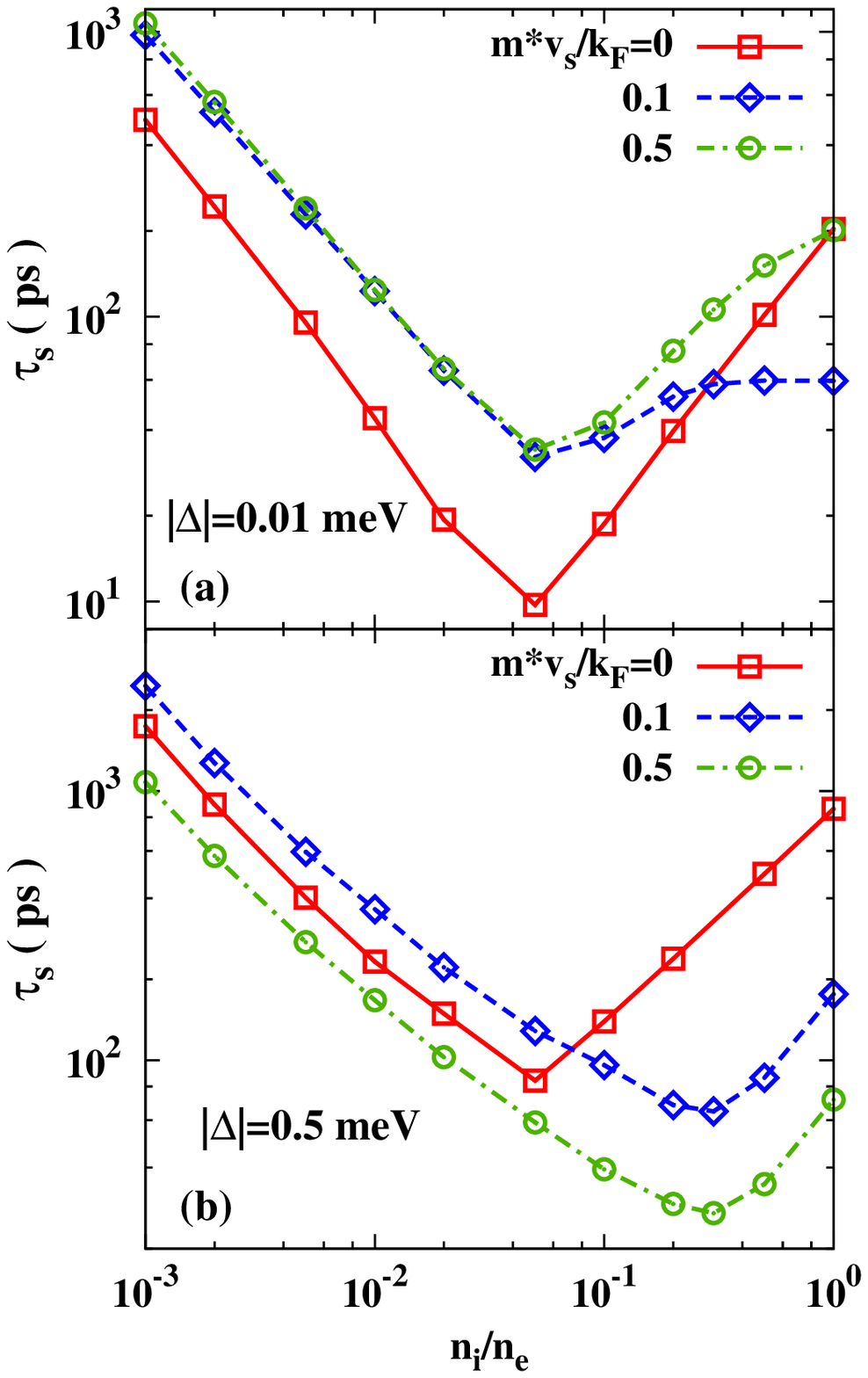}}
    \hspace{0cm}\parbox[t]{8cm}{
      \includegraphics[width=7.85cm,height=12.74cm]{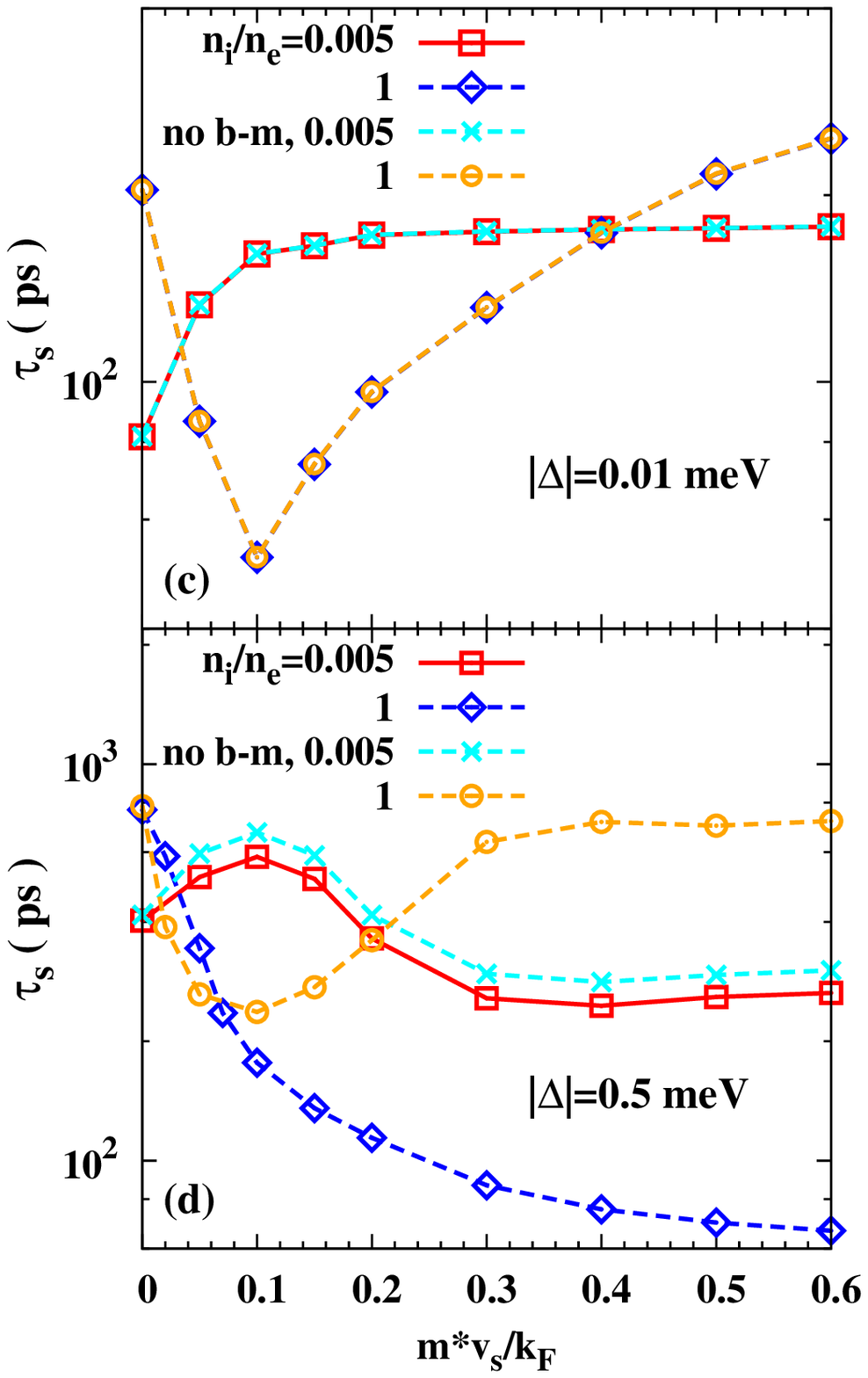}}
  \end{minipage}
\begin{minipage}[]{18cm}
\begin{center}
  \caption{(Color online) Impurity density [(a) and (b)] and superconducting
      velocity [(c) and (d)] dependencies of the SRTs. In (a) and (c) [(b) and (d)],
      $|\Delta|=0.01$ (0.5)~meV. Specifically, in (a) and (b), the SRTs are plotted  
      with different superconducting velocities $v_s=0$ (the red solid curve with
      squares), $0.1k_{\rm F}/m^*$ (the blue dashed curve with diamonds) and
      $0.5k_{\rm F}/m^*$ (the green chain curve with circles).
      In (c) and (d), the SRTs are presented with
      different impurity densities $n_i/n_e=0.005$ (the red solid curve with
      squares) and 1 (the blue dashed curve with squares). Finally, with the branch-mixing
      scattering removed, in (c) and (d), the SRTs are shown by the cyan (yellow) dashed curve with
      crosses (circles) when $n_i/n_e=0.005$ (1).}
  \label{figyw9}
\end{center}
\end{minipage}
\end{figure}
\end{widetext}

Specifically, 
in Figs.~\ref{figyw9}(a) and (b), the impurity density dependencies of the SRTs are plotted  
with different superconducting velocities $v_s=0$ (the red solid curve with
squares),
$0.1k_{\rm F}/m^*$ (the blue dashed curve with diamonds) and $0.5k_{\rm
  F}/m^*$ (the green chain curve  with circles). In Fig.~\ref{figyw9}(a), with an extremely small order parameter
$|\Delta|=0.01$~meV, it can be seen that in the weak (strong) scattering limit, the
SRT can be enhanced (suppressed) by the superconducting
velocity when $v_s\lesssim 0.5k_{\rm F}/m^*$. Moreover,  
with the increase of the superconducting velocity,  the boundary between the
weak and strong scattering regimes remains unchanged. Anomalously, when the
system enters into the strong scattering regime, for 
$v_s\gtrsim 0.1k_{\rm F}/m^*$,
the SRT becomes {\em insensitive} to the momentum scattering. However, when the order
parameter is large enough with $|\Delta|\gtrsim 0.1$~meV, the behaviors of
the SRT become different. Specifically, it can be seen
from Fig.~\ref{figyw9}(b) that when
$|\Delta|=0.5$~meV, in the weak (strong) scattering regime, the
SRT can be either enhanced or suppressed (suppressed) by the superconducting
velocity. In particular, for the superconducting velocity $v_s=0.1k_{\rm F}/m^*$, 
the boundary between the weak and strong scattering regimes is
shifted, which arises at the larger impurity density and remains unchanged with the
further increase of the superconducting
velocity (e.g., $v_s=0.5k_{\rm F}/m^*$).

To see the
role of the superconducting velocity on the quasiparticle spin relaxation more clearly, we further
calculate the superconducting velocity dependencies of the SRTs with
different impurity densities, shown in Figs.~\ref{figyw9}(c) and (d)
at low and
high  impurity densities with $|\Delta|=0.01$ and 0.5~meV.
Specifically, in Fig.~\ref{figyw9}(c), one finds that when the order parameter
is extremely small ($|\Delta|=0.01$~meV), with the increase of the
superconducting velocity, in the weak
scattering regime with $n_i/n_e=0.005$, the SRT first increases and then becomes
insensitive
to the superconducting velocity; whereas in the strong scattering regime, the 
SRT first decreases rapidly and
then increases. However, when the order parameter becomes larger (with
$|\Delta|\gtrsim 0.1$~meV), the response of
the SRTs to the superconducting velocity also becomes very
different. Specifically, it can be
seen from 
Fig.~\ref{figyw9}(d) that when $|\Delta|=0.5$~meV, with the increase of the superconducting velocity,
the SRT in the weak scattering regime ($n_i/n_e=0.005$) first increases, then decreases and finally becomes
insensitive
to the superconducting velocity; whereas in the strong scattering regime ($n_i/n_e=1$), the
SRT decreases monotonically.

Before detailed explanation for the 
rich behaviors of the quasiparticle spin relaxation in the impurity density and
superconducting velocity dependencies, we first address two important influences
of the superconducting velocity on the quasiparticle state. On one hand, when there exists
the supercurrent,
as addressed in Sec.~\ref{distribution and correlation} 
that the quasiparticle energy spectrum is tilted by the superconducting
velocity, and hence the blocking region can
appear.
Around the blocking region, the quasiparticle Fermi surface emerges, which is constituted
by the electron- and hole-like Fermi arcs (refer to
Figs.~\ref{figyw1} and \ref{figyw2}). 
On the other hand, only when $|\Delta|\gtrsim 0.1$~meV, due to the
superconducting velocity, the branch-mixing
scattering\cite{Tinkham1,Tinkham2} can be {\em efficiently} triggered due
to the tilt
of the energy spectrum (refer to Fig.~\ref{figyw2}). When $|\Delta|\rightarrow
0$, the branch-mixing scattering is forbidden due to the coherence
prefactor ($\approx \epsilon_{\bf k}/|\epsilon_{\bf k}|+\epsilon_{{\bf k}'}/|\epsilon_{{\bf
      k}'}|$) tends to zero in the momentum scattering [Eq.~(\ref{effective_U})].

\paragraph{Quasiparticle spin relaxation in a Fermi arc}
We first analyze the quasiparticle spin relaxation with extremely small
  order parameters ($|\Delta|\ll 0.1$~meV). As addressed above, the branch-mixing scattering for the
quasiparticle is forbidden, indicating that the electron- and hole-like
Fermi arcs are independent when the condensation process is slow enough.
 When $v_s\gtrsim 0.1k_{\rm F}/m^*$,
the quasiparticle distribution enters into the degenerate
  regime (refer
  to Fig.~\ref{figyw6}), and the thermal excitations of the quasiparticles happen around
  the Fermi arcs. In this situation, the quasiparticle spin relaxation can be
  simply understood by only analyzing the Fermi arc.  
Specifically,
the angular-average of the effective magnetic field in one Fermi arc, i.e., $\langle{\bgreek
  \Omega}_{\bf k}\rangle_a$, is not zero. By writing the effective magnetic field
as ${\bgreek \Omega}_{\bf k}=({\bgreek \Omega}_{\bf k}-\langle{\bgreek
  \Omega}_{\bf k}\rangle_a)+\langle{\bgreek
  \Omega}_{\bf k}\rangle_a\equiv {\bgreek \Omega}_{\bf k}^{\rm eff}+\langle{\bgreek
  \Omega}_{\bf k}\rangle_a$, one finds that around one Fermi arc, ${\bgreek \Omega}_{\bf
k}^{\rm eff}$ plays the role of the effective magnetic field
(inhomogeneous broadening\cite{wureview,inhomogeneous}) and $\langle{\bgreek
  \Omega}_{\bf k}\rangle_a$ acts as an effective Zeeman field.
Specifically, at low temperature, the effective Zeeman field can cause the 
spin oscillations even in the strong scattering regime (refer to
Appendix~\ref{CC}).

Moreover, when $|\langle{\bgreek
  \Omega}_{\bf k}\rangle_a|$ is comparable to $|{\bgreek \Omega}_{\bf
  k}^{\rm eff}|$, $\langle{\bgreek
  \Omega}_{\bf k}\rangle_a$ plays marginal role on the spin relaxation, whereas ${\bgreek \Omega}_{\bf
  k}^{\rm eff}$ leads to the spin relaxation.\cite{wureview,Zutic,Yu_Zeeman} In particular, one
finds that the magnitude of ${\bgreek \Omega}_{\bf
  k}^{\rm eff}$ {\em strongly} depends on the direction of the
momentum. Accordingly, around one Fermi arc, the variation of momentum direction
can cause the variations of both the direction and magnitude of the effective
magnetic field,
acting as the angle-dependent and modular-dependent inhomogeneous
broadenings, respectively. In this situation, even due to the elastic
scattering, the module-dependent inhomogeneous
broadening is triggered,\cite{Yu_valley} which is enhanced by the momentum
scattering.  Nevertheless, the motional narrowing effect can suppress the spin relaxation in the
strong scattering regime.\cite{DP} Thus, the competition of the
two trends leads to the insensitiveness of momentum scattering dependence of the SRT in the
strong scattering regime. The above features for the spin relaxation in the
strong scattering regime actually provide a method for experimentally
verifying the existence of the Fermi arc.

The Fermi arc structure also influences the quasiparticle spin relaxation
  in the weak scattering regime. By assuming that the proportions of the spin polarization
carried by the electron- and hole-like branches are $P_e$ and $1-P_e$,
respectively.
Then, due to the separations of the
electron-like (hole-like) Fermi arc, the angular-averaged effective
magnetic field magnitude is estimated to be $\langle|{\bgreek
  \Omega}^{\rm eff}_{\bf k}|\rangle_a \approx P_e|{\bgreek
  \Omega}_{\bf k}|$ [$(1-P_e)|{\bgreek
  \Omega}_{\bf k}|$], and the momentum scattering time becomes
$\tau_{\bf k}'\approx\tau_{\bf k}^N/P_e$ [$\tau_{\bf k}^N/(1-P_e)$]. Obviously,
$\langle |{\bgreek
  \Omega}^{\rm eff}_{\bf k}|\rangle_a \tau'_{\bf k}\approx |{\bgreek
  \Omega}_{\bf k}|\tau^N_{\bf k}$, and hence the boundary between the
weak and strong scattering regimes remains unchanged in the presence of
the supercurrent. Moreover, with $P_e$ around 1/2, the angular-average of the momentum scattering time can be
estimated by
$\langle\tau_{\bf k}'\rangle=P_e(\tau_{\bf k}^N/P_e)+(1-P_e)\tau_{\bf
  k}^N/(1-P_e)=2\tau_{\bf k}^N$. 
Then when $v_s\gtrsim 0.1k_{\rm
  F}/m^*$, the SRT in the weak scattering regime is
enhanced to be 
two times of the one with $v_s=0$, which  agrees with the
numerical calculation in Figs.~\ref{figyw9} (a) and (c) fairly well. 
It is emphasized that these
estimations are based on the Fermi arcs, which are only established when
$v_s\gtrsim 0.1k_{\rm F}/m^*$.

 Facilitated with these understandings, we then analyze the superconducting velocity dependencies of the SRT, 
explicitly shown in
Fig.~\ref{figyw9}(c).  When the superconducting velocity is small
($v_s\lesssim 0.1k_{\rm F}/m^*$), there still exists considerable quasiparticle
population in the region with $k_x>0$. In this situation, the momentum scattering
time increases from $\tau_{\bf k}^N$ to $2\tau_{\bf k}^N$ with the increase of
the superconducting velocity. Hence, in the weak (strong) scattering regime, the
SRT increases (decreases) with the increase of
the superconducting velocity. Whereas when $v_s\gtrsim 0.1k_{\rm F}/m^*$, the
physics can be simply understood based on the Fermi arc as
addressed. Accordingly,
in the weak scattering regime, the SRT 
becomes two times of the one with $v_s=0$, and remains unchanged with the
increase of the superconducting velocity. However, in
the strong scattering regime, with the increase of the superconducting velocity,
$\langle {\bgreek \Omega}_{\bf k}\rangle_a$ decreases, and the
module-dependent inhomogeneous
broadening is suppressed. Hence, with the momentum scattering time relatively unchanged, the SRT in
the strong scattering regime increases with the increase of the superconducting
velocity.

\paragraph{Branch-mixing--induced spin relaxation channel}

We then focus on the quasiparticle spin relaxation with finite order
  parameter ($|\Delta|\gtrsim 0.1$~meV). We first explain the shift of the boundary between the weak and strong
scattering regimes due to the superconducting velocity 
when $|\Delta|=0.5$~meV [Fig.~\ref{figyw9}(b)]. 
It is noticed that when $|\Delta|=0.5$~meV with $v_s=0$, from Eq.~(\ref{SOC_eff}), one
  finds that the SOC around the Fermi momentum is markedly suppressed for the
  quasiparticles.
Due to the
supercurrent, before the blocking region appears ($v_s\lesssim 0.14k_{\rm
  F}/m^*$), the quasiparticle populations are shifted away from the Fermi
momentum. This enhances not only the inhomogeneous broadening but also the momentum scattering
time due to the suppression of
the phase space available for the
quasiparticle scattering.
Thus, when $v_s=0.1k_{\rm
  F}/m^*$, these two responses to the supercurrent jointly lead
to the shift of the boundary between the weak and strong scattering regimes to a
larger impurity density. Nevertheless, once the
blocking region forms (e.g, $v_s=0.5k_F/m^*$), 
the influence of the superconducting velocity on the boundary between the
weak and strong scattering regime is similar to the
case with extremely small order parameter addressed above. Thus, when $v_s=0.5k_{\rm
  F}/m^*$ in Fig.~\ref{figyw9}(b), the
boundary between the weak and strong scattering regimes remains unchanged.

We then analyze the superconducting velocity dependencies of the
SRTs, shown in Fig.~\ref{figyw9}(d). Unlike the situation with an
  extremely small order parameter, the branch-mixing scattering can be
efficiently triggered in the presence of the blocking region. Here,
the role of the branch-mixing scattering on the quasiparticle
spin relaxation is revealed by comparing the SRTs with and without the
branch-mixing scattering.  In Fig.~\ref{figyw9}(d), with the branch-mixing
scattering switched off, in the strong (weak) scattering regime,
it is shown by the yellow (cyan) dashed curve with circles (crosses) that the
SRTs are much larger than (comparable to) the ones with the branch-mixing
scattering when $v_s\gtrsim
0.14 k_{\rm F}/m^*$. Hence, the branch-mixing
scattering plays dominant (marginal) role on the quasiparticle spin relaxation
in the strong (weak) scattering regime in the presence of the blocking region.
It is noted that the spin relaxation channel due to the branch-mixing scattering 
resembles the one due to the inter-valley scattering shown in our previous
studies on the
electron spin relaxations in monolayer rippled\cite{rippled_graphene} and
bilayer\cite{bilayer graphene}
graphene or monolayer MoS$_2$.\cite{monolayer_MoS2}

Finally, we address a new feature in the weak scattering regime when
$|\Delta|=0.5$~meV compared to the case with extremely small order parameter
$|\Delta|=0.01$~meV. In Fig.~\ref{figyw9}(d), one notices that when
$|\Delta|=0.5$~meV with $n_i/n_e=0.005$, with the increase of the superconducting velocity in
the region $0.1k_{\rm
  F}/m^*\lesssim v_s\lesssim 0.3k_{\rm
  F}/m^*$, the SRT 
is suppressed. This suppression of the SRT comes from the quasiparticle density
dependencies on the superconducting velocity (refer to
Fig.~\ref{figyw6}).  
When $0.1k_{\rm
  F}/m^*\lesssim v_s\lesssim 0.3k_{\rm
  F}/m^*$, the system lies in the crossover between the
non-degenerate and degenerate limits. As predicted in the Brooks-Herring
formula,\cite{Brooks_Herring} in the non-degenerate regime, 
the momentum scattering due to impurities is enhanced with the increase of the quasiparticle density, owing to 
the increase of the phase space
available for the quasiparticle scattering; whereas in the degenerate regime, the
quasiparticle-impurity scattering becomes insensitive to the quasiparticle
density.\cite{Zutic,wureview} Hence, when $v_s\gtrsim 0.1k_{\rm F}/m^*$, the SRT first decreases and
then becomes unchanged with the increase of the
superconducting velocity.

\section{Conclusion and discussion}
\label{summary}
In conclusion, we have investigated the DP spin relaxation for the quasiparticle
with a novel superconducting-velocity--tunable state in GaAs
(100) QWs in proximity to an $s$-wave superconductor, by the KSBEs derived in the
quasiparticle approximation. 
 In the superconducting QW, the superconducting velocity induced from the
 superconducting phase is shown to cause the tilt
 of the quasiparticle energy
spectrum. It is found that when the
quasiparticle energy spectrum is tilted to be smaller than the chemical
potential, a blocking
region corresponding to
the  negative
excitation energy appears, which is in the crescent shape. 
The existence of the blocking region can significantly influence
the Cooper pairings, quasiparticle density and momentum current
in QWs.

Specifically, the
center-of-mass momentum ${\bf q}$ is carried by the Cooper pairs, with the
anomalous correlations only existing between the states with momentum ${\bf k}+{\bf q}/2$
and $-{\bf k}+{\bf q}/2$. Moreover, the
Cooper pairings around the electron Fermi surface are efficiently suppressed. Furthermore,
the quasiparticle density increases with the increase of the superconducting
velocity. Particularly, the degenerate regime can be realized and the Fermi
  surface around the blocking region for the quasiparticles
  appears. The quasiparticle Fermi surface is constituted by the electron- and hole-like
  Fermi arcs. Finally, it is revealed that when the blocking region
appears, the momentum current contributed by the quasiparticles flows in the opposite
direction to the one
carried by the Cooper
pairs. Thus, in the superconducting
velocity dependence of the momentum current, a peak arises due to the competition of 
the Cooper pairs and quasiparticles in the blocking region, whose appearance  
corresponds to the emergence of the blocking region.

We then study the DP spin relaxation for the quasiparticles in the
  superconducting QWs by the KSBEs.  The KSBEs for the quasiparticle is set
up based on the quasiparticle approximation, in which the SOC and quasiparticle-impurity
scattering are explicitly considered. By using the KSBEs, the SRTs without and with the
supercurrent are calculated. Rich physics is
revealed. Without the supercurrent, 
we address that the branch-mixing scattering\cite{Tinkham1,Tinkham2} due to
  the impurity is forbidden, indicating that the electron- and hole-like
  branches are independent and hence only the intra-branch spin relaxation
    channel exists. In this
situation, when $|\Delta|$ tends to zero, the SRT recovers to the normal one.
 Whereas with finite order parameter, we find that compared to the normal state, both the
SOC and quasiparticle-impurity scattering are efficiently suppressed due to the
{\it same} coherence factor $|\epsilon_{\bf k}|/\mathcal{E}_{\bf k}$. This leads to
the enhancement of the SRT in both weak and strong scattering regimes with
the boundary between the weak and
strong scattering regimes unchanged.

When the supercurrent is turned on, the Fermi arc has intriguing effect on the DP spin relaxation,
  which exhibits very different physics depending on the magnitude of the order
  parameter.  Specifically, when the order parameter is extremely small (e.g.,
$|\Delta|=0.01$~meV), the branch-mixing scattering is still forbidden because the coherence factor
($\approx\epsilon_{\bf k}/|\epsilon_{\bf k}|+\epsilon_{{\bf k}'}/|\epsilon_{{\bf k}'}|$) tends to be
zero in the quasiparticle-impurity scattering, indicating that the electron- and hole-like
Fermi arcs are independent. 
Anomalous features are revealed for the quasiparticle spin relaxation at
the Fermi arc in the {\em strong} scattering
regime. Specifically, on one hand, the spin oscillations can be induced by the superconducting
velocity; on the other hand, the SRT becomes insensitive to the momentum scattering.
The latter feature is in contrast to the
motional narrowing feature in the conventional DP relaxation,\cite{DP}
in which the spin relaxation is suppressed by the
momentum scattering.

With the {\em
  finite} order parameter ($|\Delta| \gtrsim 0.1$~meV), the situation becomes
different. It is revealed that the tilt
of the energy spectrum can trigger the branch-mixing scattering. 
In this situation,
there exist not only the intra-branch spin relaxation channel, but also the
inter-branch one.
Furthermore, we reveal the role of the intra- and inter-branch spin
relaxation channels on the spin relaxation in both the weak and strong
scattering regimes. Specifically, in the weak scattering regime, the intra-branch spin relaxation
channel is dominant; whereas in the strong scattering regime, the inter-branch
channel becomes dominant when the blocking region appears. Moreover, it is
  found that in contrast to the situation with an extremely small order parameter, in the
strong scattering regime, no spin oscillation occurs and the spin relaxation
is suppressed by the momentum scattering.

Up till now, what predicted in this investigation has not yet been
experimentally reported and we expect that our work will bring
experimental attention. For the novel superconducting-velocity-tunable
quasiparticle state in the QWs, the suppressed Cooper pairings,
large quasiparticle density and non-monotonically tunable
momentum current can be directly measured, which reflects the information of the
tilted quasiparticle energy spectrum and the emergence of the blocking
region. For the quasiparticle spin relaxation, when the order parameter is
extremely small as reported in Ref.~\onlinecite{2D_super_GaAs}
($|\Delta|=46$~$\mu$eV), in the presence of the supercurrent, the spin oscillations and the
insensitiveness of the SRT on the impurity density can provide the evidence for the existence of the
Fermi arcs; when the order parameter is finite, in the strong scattering regime,
the absence of the spin oscillations and the suppression of the SRT by
 the relatively large superconducting velocity (larger than the critical
 velocity) show the effect of the branch-mixing
scattering on the quasiparticle spin relaxation.

Finally, we point out that the existence of the Fermi arcs can be markedly
influenced by the dynamics of the quasiparticle condensation. It is important 
that the condensation rate is slower than the spin relaxation
one. However, up till now, the details of the condensation process are not
clear in superconductors,\cite{Bardeen,Josephson,Tinkham_book} not to mention in the system proximity to
a superconductor. Therefore, the study on the spin dynamics and the
existence of the Fermi arc can even shed light on the
condensation process.

\begin{appendix}

\section{SPIN OSCILLATIONS INDUCED BY SUPERCONDUCTING VELOCITY}
\label{CC}
In this appendix, we show the spin oscillations induced by the superconducting
velocity in the strong scattering regime with extremely small order
parameter. It has been addressed in Sec.~\ref{with_superfluid} that the angular-average of the
effective magnetic field in one Fermi arc is not zero, which acts as an
effective Zeeman field. Accordingly, even in the strong scattering regime, this
effective Zeeman field can lead to the spin oscillations.

\begin{figure}[ht]
  {\includegraphics[width=8.5cm]{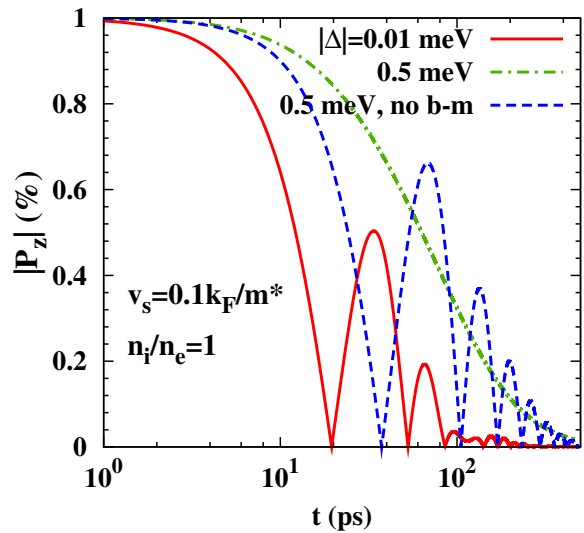}}
  \caption{(Color online) Temporal evolution of the spin polarization in the strong
    scattering regime with
    extremely small order parameter $|\Delta|=0.01$~meV (the red solid curve) and finite order
    parameter $|\Delta|=0.5$~meV (the green chain curve). $v_s=0.1k_F/m^*$ The impurity density
    is $n_i/n_e=1$, indicating that the system lies in the strong scattering
    regime. The blue dashed curve represents the situation by switching off the branch-mixing scattering.}
  \label{figyw11}
\end{figure}

In Fig.~\ref{figyw11}, it is shown by the red solid curve that when the order
parameter is extremely small ($|\Delta|=0.01$~meV), in the strong scattering
regime ($n_i/n_e=1$) with the supercurrent ($v_s=0.1k_F/m^*$), the spin
polarization oscillates in the temporal evolution.
Whereas with a finite order
parameter $|\Delta|=0.5$~meV, it is shown by the green chain curve that the
oscillations vanish. This is
because in the strong scattering regime here, the triggered branch-mixing
scattering becomes important. Thus, the quasi-electron can feel the SOC
around the Fermi surface, whose angular-average is exactly zero. By arbitrarily
switching off the branch-mixing scattering, the spin oscillations of the spin
polarization occur, shown by the blue dashed curve.

\end{appendix}

\begin{acknowledgments}
This work was supported
 by the National Natural Science Foundation of China under Grant
No. 11334014 and  61411136001, the National Basic Research Program
 of China under Grant No.
2012CB922002 and the Strategic Priority Research Program 
of the Chinese Academy of Sciences under Grant
No. XDB01000000.

\end{acknowledgments}

\end{document}